\documentclass[twocolumn,tighten,times,astrosymb]{aastex631}

\usepackage{float}
\usepackage{amsmath}
\usepackage{microtype}
\hypersetup{linkcolor=red,citecolor=green,filecolor=cyan,urlcolor=magenta,pdftitle={Thermodynamical Description of Hot, Rapidly Rotating Neutron Stars, Protoneutron Stars, and Neutron Star Merger Remnants}}
\usepackage{booktabs}
\usepackage{hyperref}

\received{July 20, 2020}
\revised{January 29, 2021}
\accepted{February 10, 2021}

\submitjournal{ApJ}

\shortauthors{Koliogiannis \& Moustakidis}

\begin{document}
	
	\title{Thermodynamical Description of Hot, Rapidly Rotating Neutron Stars, Protoneutron Stars, and Neutron Star Merger Remnants}

	\correspondingauthor{P.S. Koliogiannis}
	\email{pkoliogi@physics.auth.gr}
	
	\author[0000-0001-9326-7481]{P.S. Koliogiannis}
	\affiliation{Department of Theoretical Physics, Aristotle University of Thessaloniki, 54124 Thessaloniki, Greece}
	
	\author[0000-0003-3380-5131]{Ch.C. Moustakidis}
	\affiliation{Department of Theoretical Physics, Aristotle University of Thessaloniki, 54124 Thessaloniki, Greece}
	
	\begin{abstract}
		The prediction of the equation of state of hot, dense nuclear matter is one of the most complicated and interesting problems in nuclear astrophysics. At the same time, knowledge of it is the basic ingredient for some of the most interesting studies. In the present work, we concentrate our study on the construction of the equation of state of hot, dense nuclear matter, related mainly to the interior of the neutron star. We employ a theoretical nuclear model, which includes momentum-dependent interaction among the nucleons, along with the \textit{state-of-the-art} microscopic calculations. Thermal effects are introduced in a self-consistent way, and a set of isothermal and isentropic equations of state are predicted. The predicted equations of state are used in order to acquire and to extend the knowledge of the thermal effect on both nonrotating and rapidly rotating with the Kepler frequency neutron stars. The simultaneous study of thermal and rotation effects provides useful information on some of the most important quantities, including the mass (gravitational and baryon) and radius, the Kepler frequency and Kerr parameter, the moment of inertia, etc. These quantities are directly related to studies of protoneutron stars and mainly the hot and rapidly rotating remnant of a binary neutron star merger. Data from the late observations of binary neutron star mergers and the present study may offer useful tools for their investigation and help in providing possible constraints on the equation of state of nuclear matter. 
	\end{abstract}
	
	
	\section{Introduction}
	Neutron stars are a way for the universe to manifest its densest objects with an internal structure. Their study requires the use of general relativity, as their hydrodynamical equilibrium is described by the Tolman-Oppenheimer-Volkoff (TOV) differential equations. In order to proceed with the TOV equations, the equation of state (EOS) of the fluid in the interior of the star is mandatory. However, the EOS still remains uncertain from both a theoretical and an experimental point of view. This uncertainty is well reflected on the predicted bulk properties of a neutron star. Since there is a limitation on the experimental data concerning the dense nuclear matter, we concentrate on the systematic study of the existing observational ones. These studies are mainly related to the observation of isolated nonrotating or rotating neutron stars and the evolution of pulsars, as well as of binary neutron stars in neutron star-black hole systems, supernova explosions, etc.
	
	Until this moment, the observation of non/slow-rotating neutron stars has provided us with severe constraints on the dense nuclear matter through their maximum possible mass. The most recent ones are the PSR J0740+6620 with $M=2.14_{-0.09}^{+0.10}~M_{\odot}$ (Cromartie et al.~\citeyear{Cromartie-2019}) and the more uncertain PSR J2215+5135 with $M=2.27_{-0.15}^{+0.17}~M_{\odot}$ (Linares et al.~\citeyear{Linares-2018}). However, the recent observation of gravitational waves from a merging neutron star binary system (GW170817; Abbott et al.~\citeyear{PhysRevLett.119.161101}) opened a new, very important source to probe and improve our knowledge of the EOS in multiple ways. To be more specific, the EOS of both cold and hot nuclear matter considerably affects the dynamic process of the prior and postmerger phase of binary neutron stars, which lead to a hot remnant. This process also includes the tidal polarizability during the inspiral of a binary system. In addition, after the merger, the maximum stable mass, the spin period, and the lifetime of the remnant strongly depend on the dense matter properties at high temperature and entropy. In particular, the evolution and possible final stage of the remnant are sensitive to the EOS, including (a) the time scale for the gravitational collapse to a black hole; (b) the possibility of a phase transition to other degrees of freedom (hyperons, quarks, etc.), which may lead to collapse to a black hole (due to softening of the EOS); and (c) the creation of a disk around the remnant, ejecta, and neutrino emission.
	
	In earlier years, pioneering work was done for the study of a hot EOS for astrophysical applications, including the studies of Bethe et al.~\citeyearpar{BETHE1979487}, Brown et al.~\citeyearpar{BROWN1982481}, Lamb et al.~\citeyearpar{PhysRevLett.41.1623}, Lattimer \& Ravenhall~\citeyearpar{As.J.223.314} and Lattimer~\citeyearpar{AnnRNPS.31.337}. Over the years, the most used EOSs of hot neutron star matter have been (a) the liquid drop-type model constructed by Lattimer \& Swesty~\citeyearpar{LATTIMER1991331} and (b) the one by Shen et al.~\citeyearpar{SHEN1998435}, where the relativistic mean field model is employed. Later on, Shen et al.~\citeyearpar{SHEN1998435} extended their study to generate EOSs of nuclear matter for a wide range of temperatures, densities, and proton fractions for applications in supernovae, neutron star mergers, and black hole formation simulations by also employing a full relativistic mean field (Shen et al.~\citeyear{PhysRevC.83.065808}).
	
	Wellenhofer et al.~\citeyearpar{PhysRevC.92.015801} investigated  the density and temperature dependence of the nuclear symmetry free energy using microscopic two- and three-body nuclear potentials constructed from Chiral effective field theory. Constantinou et al.~\citeyearpar{PhysRevC.89.065802,PhysRevC.92.025801} derived a hot EOS suitable to describe supernova and hot neutron star properties. Temperature effects on the neutron star matter EOS were investigated in the framework of Chiral effective field theory by Sammarruca et al.~\citeyearpar{doi:10.1142/S0217732320501564}. The properties of hot $\beta-$stable nuclear matter, using EOSs derived within the Brueckner-Hartree-Fock approach at finite temperature, have been provided in a series of papers (Nicotra et al.~\citeyear{AA.451.1.2010}; Burgio \& Schulze~\citeyear{AA.518.A17.2010}; Baldo \& Burgio~\citeyear{BALDO2016203}; Fortin et al.~\citeyear{10.1093/mnras/sty147}; Lu et al.~\citeyear{PhysRevC.100.054335},~\citeyear{PhysRevC.103.024307}; Figura et al.~\citeyear{PhysRevD.102.043006}; Shang et al.~\citeyear{PhysRevC.101.065801}; Wei et al.~\citeyear{10.1093/mnras/staa1879}). Raithel et al.~\citeyearpar{Raithel_2019} derived a model that allows the extension of any cold nucleonic EOS, including piecewise polytropes, to arbitrary temperature and proton fractions for use in calculations and numerical simulations of astrophysical phenomena.
	
	Moreover, a detailed study of the evolution of protoneutron stars was predicted by Pons et al.~\citeyearpar{Pons_1999} and Prakash et al.~\citeyearpar{Prakash_2000}. The authors focused on the thermal and chemical evolution of the birth of neutron stars by employing neutrino opacities consistently calculated with the underlying nuclear EOS (Pons et al.~\citeyear{Pons_1999}). For a recent review of the hot EOS of dense matter and neutron stars, see Lattimer \& Prakash~\citeyearpar{LATTIMER2016127}.
	
	In the last 40 yr, a lot of theoretical work has been dedicated to studying the processes of the merger and postmerger phases of a binary neutron star system, and important progress has been achieved. However, there are many relevant issues that remain unsolved, or at least under consideration. In general, we refer to the remnant evolution, mainly including the collapse time and threshold mass. Moreover, the possibility of a phase transition in the interior of the remnant may affect the signal of the emitted gravitational waves. In addition, matters under consideration are also the disk ejecta and neutrino emission properties, which are sensitive to the employed EOS (for an extended discussion and applications, see Perego et al.~\citeyear{Perego2019}). Some previous work is also included in Bauswein et al.~\citeyearpar{PhysRevD.82.084043}, Kaplan et al.~\citeyearpar{Kaplan_2014}, Tsokaros et al.~\citeyearpar{PhysRevLett.124.071101}, Yasin et al.~\citeyearpar{PhysRevLett.124.092701}, Radice et al.~\citeyearpar{doi:10.1146/annurev-nucl-013120-114541}, Sarin et al.~\citeyearpar{PhysRevD.101.063021}, Soma \& Bandyopadhyay~\citeyearpar{Soma_2020}, and Sen~\citeyearpar{Sen_2020}.
	
	It is worth mentioning that the theory of quantum chromodynamics (QCD) predicts the ongoing transition of hadron matter to unconfined quark matter at a sufficiently high density (a few times the saturation density). As neutron stars provide a rich testing ground for microscopic theories of dense nuclear matter, combining this study with the experimental data from ultrarelativistic heavy-ion collisions (the Relativistic Heavy Ion Collider at Brookhaven and the Large Hadron Collider at CERN) may help to significantly improve our knowledge of QCD theory (Baym et al.~\citeyear{Baym_2018}). However, the problem of the existence of free quark matter in the interior of neutron stars remains. Moreover, the emergence of strange hadrons (hyperons, etc.) around twice the nuclear saturation density leads to an appreciable softness of the EOS and low values of neutron star mass, far from observation. This problem is highlighted as the hyperon puzzle. Of course, there are other studies where the authors stated that hyperon consideration on the EOS is not in contradiction with the predictions of a very high neutron star mass (see Chatterjee \& Vida$\rm \tilde{n}$a~\citeyear{Chatterjee_2016}; Li et al.~\citeyear{LI2020135812}).
		
	Recently, it has been claimed that the recent observation of gravitational waves from neutron star mergers could shed light on the possibility of hadrons moving to a quark phase transition (Annala et al.~\citeyear{Annala2020}). The authors stated that if the conformal limit on the value of the speed of sound, $c_{s}/c \leq 1/\sqrt{3}$, is not strongly violated, then heavy neutron stars may have sizable quark matter cores. In this case, important implications must be considered in neutron star mergers with at least one massive participant (Annala et al.~\citeyear{Annala2020}). However in the present work, we do not consider the case of additional degrees of freedom (hyperons, quarks, etc.) in the interior of neutron stars. This issue will be under consideration in a future study.
	
	The present work consists of two parts. In the first part of the paper, we investigate the bulk properties of hot nuclear and neutron star matter. In particular, we apply a momentum-dependent effective interaction (MDI) model, where thermal effects can be studied simultaneously on the kinetic part of the energy and also on the interaction one. The advantage of the present model, compared to others, is that thermal effects are introduced in a self-consistent way. To be more specific, we rigorously enforce the thermodynamic laws describing the hot dense nuclear matter. In addition, this model can be extended in order to modify the stiffness of the proposed EOS by properly parameterizing the nuclear symmetry energy. It is worth pointing out that a large number of EOSs of hot nuclear and neutron star matter for astrophysical applications have appeared over the years, employing various theoretical models and approximations. However, most of them are questionable in the sense that thermal effects are not included in the cold EOS in a self-consistent way but rather in an artificial one. This point has already been noted in Constantinou et al.~\citeyearpar{PhysRevC.92.025801}. Actually, the present model was introduced by Gale et al.~\citeyearpar{PhysRevC.35.1666} in order to examine the influence of MDI on the momentum flow of heavy-ion collisions. Nonetheless, over the years, the model has been extensively applied to study the properties of cold and hot nuclear and neutron star matter (for a review of the model, see Prakash et al.~\citeyear{PRAKASH19971}; Li \& Schr{\"o}der~\citeyear{Libook}; Li et al.~\citeyear{LI2008113}).
	
	Moving on to the second part of the paper, a set of thermodynamically consistent isothermal and isentropic EOSs, based on the parameterized cold one, are produced. Our eventual purpose is the application of the predicted EOSs for an extensive study on the bulk properties (including mainly the mass and radius, moment of inertia, Kerr parameter, etc.) both at nonrotating and rotating with the Kepler frequency neutron stars, as well as protoneutron stars, and neutron star merger remnants. We pay special attention to the sequences of constant baryon mass (baryon mass is equal to rest mass) and examine the peculiar role of the Kerr parameter. Finally, we dedicate a part for the study of a few postmerger processes, such as the hot, rapidly rotating remnant and the threshold mass, and we connect them with the derived EOSs.
	
	The paper is structured as follows. In Section~\ref{section:the nuclear model} we present the details of the theoretical nuclear model, paying special attention to the specific parameterization. In Section~\ref{section:Thermodynamical description of hot nuclear matter} a thermodynamical description of hot nuclear matter is provided, while in Section~\ref{section:Rapidly rotating hot neutron stars}, the rapidly rotating configuration is analyzed. Section~\ref{section:Discussion and Conclusions} is dedicated to the discussion of the conclusions, and Section~\ref{section:Remarks} lays out the scientific remarks. Finally, Section~\ref{section: numerical code} contains the computational recipe, and the \hyperref[section:appendix_1]{Appendix} provides the properties of nuclear matter.
	
	\section{The nuclear model} \label{section:the nuclear model}
	\subsection{MDI Model}
	The MDI model, applied in the present work, combines both density and MDI among the nucleons. The main origin of the momentum dependence in the Brueckner theory is the nonlocality of the exchange interaction. It was stated by Bertsch \& Gupta~\citeyearpar{BERTSCH1988189} that a single particle potential, which depends only on the baryon density, is oversimplified. In particular, it is well known that nuclear interaction has strong exchange effects that give rise to a momentum dependence in the single particle potential, and, as a consequence, it has an effect on the energy density functional. The present model was introduced by Gale et al.~\citeyearpar{PhysRevC.35.1666}, Gale et al.~\citeyearpar{PhysRevC.41.1545}, Bertsch \& Gupta~\citeyearpar{BERTSCH1988189}, and Prakash et al.~\citeyearpar{PhysRevC.37.2253} to examine the influence of MDIs on the momentum flow of heavy-ion collisions. Over the years, the model has been modified, elaborated, and extensively applied in the study of not only heavy-ion collisions but also the properties of nuclear matter (Csernai et al.~\citeyear{PhysRevC.46.736}; Sumiyoshi \& Toki~\citeyear{Ap.J.422/700}; Modarres~\citeyear{Modarres_1997}; Das et al.~\citeyear{PhysRevC.67.034611},~\citeyear{PhysRevC.75.015807}; Li et al.~\citeyear{PhysRevC.69.011603}, ~\citeyear{LI2004563}; Chen et al.~\citeyear{PhysRevLett.94.032701}; Xu et al.~\citeyear{PhysRevC.75.014607}). In the following, we present some details of the model.
	
	The energy density of the asymmetric nuclear matter is given by the relation
	\begin{equation}
		\mathcal{E}(n_n,n_p,T)=\mathcal{E}_{\rm kin}^{n}(n_n,T)+\mathcal{E}_{\rm kin}^{p}(n_p,T)+
		V_{\rm int}(n_n,n_p,T),
		\label{eq_s2:en_d}
	\end{equation}
	where $n_n$, $n_p$, and $n=n_n+n_p$ are the neutron, proton, and total baryon density, respectively. The specific contribution of the kinetic parts is given by the integrals  
	\begin{equation}
		\mathcal{E}_{\rm kin}^{\tau}(n_{\tau},T) = 2 \int \frac{d^3 k}{(2 \pi)^3}\frac{\hbar^2 k^2}{2m} f_{\tau}(n_{\tau},k,T),
		\label{eq_s2:en_k}
	\end{equation}
	where $\tau=n,p$ and $f_{\tau}$ is the Fermi-Dirac distribution function with the form
	\begin{align}
		f_{\tau}(n_{\tau},k,T)=\left[1+\exp\left(\frac{e_{\tau}(n_{\tau},k,T)-\mu_{\tau}(n_{\tau},T)}{T}\right)\right]^{-1},
		\label{eq_s2:fd}
	\end{align}
	with $e_{\tau}(n_{\tau},k,T)$ being the single particle energy and $\mu_{\tau}(n_{\tau},T)$ being the chemical potential for each species. As for the nucleon density $n_{\tau}$, its evaluation is possible through the integral
	\begin{equation}
		n_{\tau}=2 \int \frac{d^3k}{(2\pi)^3}f_{\tau}(n_{\tau},k,T).
		\label{eq_s2:nd}
	\end{equation}
	The single particle energy is available through the form
	\begin{equation}
		e_{\tau}(n_{\tau},k,T)=\frac{\hbar^2k^2}{2m}+U_{\tau}(n_{\tau},k,T),
		\label{eq_s2:spe}
	\end{equation}
	where the single particle potential $U_{\tau}(n_{\tau},k,T)$ is obtained by the functional derivative  of the interaction part of the energy density with respect to the distribution function $f_{\tau}$. Including the effect of finite range forces among nucleons, in order to avoid acausal behavior at high densities, the potential contribution is parameterized as follows (Prakash et al.~\citeyear{PRAKASH19971}):
	\begin{equation}
		V_{\rm int}(n_n,n_p,T) = V_{A} + V_{B} + V_{C},
		\label{eq_s2:p}
	\end{equation}
	with
	\begin{align}
		V_{A} =& \frac{1}{3}An_{s}\left[\frac{3}{2}-\left( \frac{1}{2}+x_{0} \right)I^2\right]u^2, \label{eq_s2:p_par1}\\
		V_{B} =& \frac{\frac{2}{3}Bn_{s}\left[\frac{3}{2}-\left(\frac{1}{2}+x_{3}\right)I^2\right]u^{\sigma+1}}
		{1+\frac{2}{3}B^{\prime}\left[\frac{3}{2}-\left(\frac{1}{2}+x_{3}\right)I^2\right]u^{\sigma-1}}, \label{eq_s2:p_par2} \\
		V_{C} =& u \sum_{i=1,2}\left[C_i \left(\mathcal{J}_{n}^{i}+\mathcal{J}_{p}^{i}\right) + I\frac{(C_i-8Z_i)}{5}\left(\mathcal{J}_{n}^{i}-\mathcal{J}_{p}^{i}\right)\right],
		\label{eq_s2:p_par3}
	\end{align}
	where $n_{s}$ denotes the saturation density, $u=n/n_{s}$, $I=1-2Y_{p}$ is the asymmetry parameter, $Y_{p}$ is the proton fraction, $[A,B,B^{\prime},C_{i}]$ are the parameters for symmetric nuclear matter (SNM), $[x_{0},x_{3},Z_{i}]$ are the parameters for asymmetric nuclear matter, and
	\begin{equation}
		\mathcal{J}_{\tau}^{i}= 2 \int \frac{d^3k}{(2\pi)^3}g(k,\Lambda_i)f_{\tau}(n_{\tau},k,T),
		\label{eq_s2:J}
	\end{equation}
	with $g(k,\Lambda_{i})$ being a suitable function to simulate finite range effects.
	
	\subsection{The Parameterization of the Model}
	\begin{figure}
		\includegraphics[width=\columnwidth]{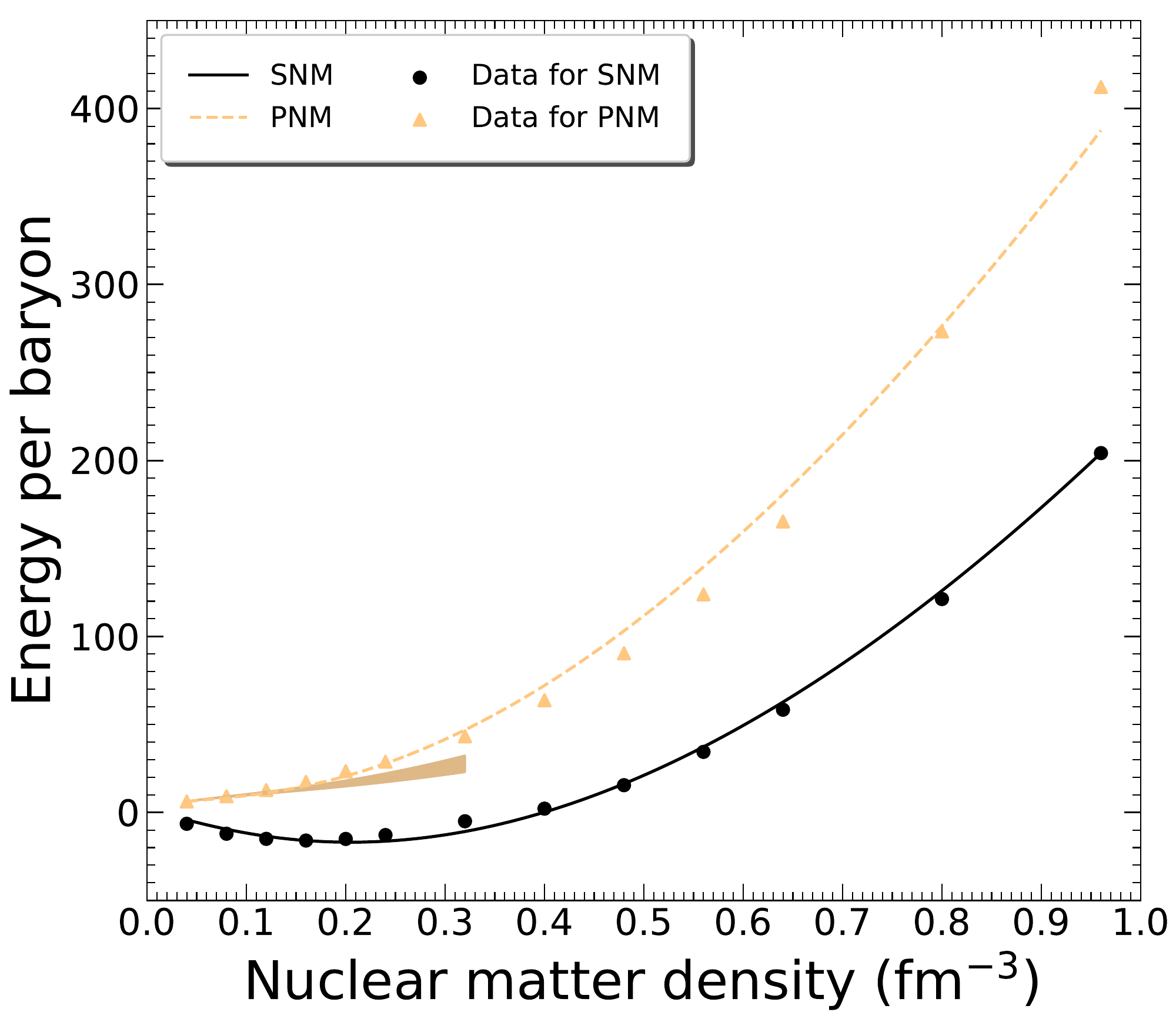}
		\caption{The SNM and PNM fits for the MDI+APR1 cold EOS. The SNM is presented by the circles and solid line, while the PNM is presented by the triangles and dashed line. The shaded region corresponds to benchmark calculations of the energy per particle of PNM extracted from Piarulli et al.~\citeyearpar{PhysRevC.101.045801}.}
		\label{fig:snm+pnm}
	\end{figure}

	The construction of the EOSs for the description of neutron stars is based on the MDI model and the data provided by Akmal et al.~\citeyearpar{PhysRevC.58.1804} for the APR-1 EOS (hereafter MDI+APR1). Its schematic presentation is shown in Figure~\ref{fig:snm+pnm}. This model, as a microscopic one, is available via ab initio calculations. The explicit use of the MDI model is due not only to its numerous advantages but also to its ability to express the energy per particle as a function of the density and momentum. This property is the one that allows the extension of its parameterization to a finite temperature that is suitable for studying processes sensitive to thermal effects, including core-collapse supernovae, protoneutron stars, neutron star mergers, etc.
	
	Using this parameterization we have constructed one cold EOS, 10 hot EOSs based on various temperatures in the range $[1,60]$ MeV, and nine hot EOSs based on various lepton fractions and entropies per baryon in the ranges $[0.2,0.4]$ and $[1,3]~k_{B}$, respectively. The advantages of the MDI+APR1 EOS are (a) it reproduces with high accuracy the properties of SNM at the saturation density (including isovector quantities $K_{s}$ and $Q_{s}$) which are shown in Table~\ref{tab:properties of nm}; (b) it correctly reproduces the microscopic calculations of the Chiral model (Hebeler \& Schwenk~\citeyear{PhysRevC.82.014314}) for pure neutron matter (PNM;for low densities) and the results of state-of-the-art calculations of Akmal et al.~(\citeyear{PhysRevC.58.1804}; for high densities); and (c) it predicts a maximum neutron star mass at least higher than the observed ones (Demorest et al.~\citeyear{Nature467}; Antoniadis et al.~\citeyear{Antoniadis-2013}; Fonseca et al.~\citeyear{Fonseca_2016}; Arzoumanian et al.~\citeyear{Arzoumanian-2018}; Linares et al.~\citeyear{Linares-2018}; Cromartie et al.~\citeyear{Cromartie-2019}). In addition, we have compared the predictions of PNM with those originating from the very recent state-of-the-art calculations (shaded region in Figure~\ref{fig:snm+pnm}; Piarulli et al.~\citeyear{PhysRevC.101.045801}). From Figure~\ref{fig:snm+pnm}, it is obvious that at very low densities, the agreement is quite satisfactory, while for higher densities, a deviation is exhibited. The latter is pointed out and discussed in Piarulli et al.~\citeyearpar{PhysRevC.101.045801}.
	
	\begin{table}
		\footnotesize
		\caption{Properties of Nuclear Matter (NM) at the Saturation Density for the MDI+APR1 EOS}
		\begin{tabular*}{\columnwidth}{@{\extracolsep{\fill} }  l  r  r }
			\toprule
			\toprule
			Properties of NM & MDI+APR1 & Units \\
			\midrule
			\decimals
			$L$ & 77.696 & MeV \\
			$Q_{\rm sym}$ & 223.061 & MeV \\
			$K_{\rm sym}$ & 0.016 & MeV \\
			$E_{\rm sym}$ & 31.071 & MeV \\
			$Q_{s}$ & -25.687 & MeV \\
			$K_{s}$ & 220.671 & MeV \\
			$m^{*}_{\tau}/m_{\tau}$ & 0.822 \\
			\bottomrule
		\end{tabular*}
		\label{tab:properties of nm}
		\tablecomments{\footnotesize Description of the reported quantities are provided in the \hyperref[section:appendix_1]{Appendix}.}
	\end{table}
	
	For the solid crust region, we adopted two models. For the cold case, we applied the EOS of Feynman et al.~\citeyearpar{PhysRev.75.1561} and also Baym et al.~\citeyearpar{Ap.J.170/299}, while for the finite temperature cases and the low-density region $(n_{b}\leq 0.08~{\rm fm^{-3}})$, as well as the finite entropies per baryon and lepton fractions, the EOSs of Lattimer \& Swesty (Lattimer \& Swesty~\citeyear{LATTIMER1991331}; hereafter LS220) and the specific model corresponding to the incomprehensibility modulus at the saturation density of SNM $K_{s}=220~{\rm MeV}$ are used (\url{https://www.stellarcollapse.org}).
	
	\section{Thermodynamics of hot neutron star matter} \label{section:Thermodynamical description of hot nuclear matter}
	The study of the properties of nuclear matter at finite temperatures requires the knowledge of the Helmholtz free energy $F$. The differentials of the total free energy $F_{\rm tot}$ and the total internal energy $E_{\rm tot}$ (total free/internal energy of baryons contained in volume $V$) are given as (Goodstein~\citeyear{Goodstein-85}; Fetter \& Walecka~\citeyear{Fetter-03})
	\begin{align}
		dF_{\rm tot} &= -S_{\rm tot}dT-PdV+\sum_i \mu_i dN_i, \label{eq_s3:Ftot} \\
		dE_{\rm tot} &= TdS_{\rm tot}-PdV+\sum_i \mu_i dN_i,
		\label{eq_s3:Etot}
	\end{align}
	where $S_{\rm tot}$ is the total entropy of baryons, and $\mu_i$ and $N_i$ are the chemical potential and number of particles of each species, respectively. The free energy per particle $F$ can be written as
	\begin{equation}
		F(n,T,I)=E(n,T,I)-TS(n,T,I),
		\label{eq_s3:fe}
	\end{equation}
	with $E=\mathcal{E}/n$ and $S=s/n$ being the internal energy and entropy per particle, respectively. It has to be noted here that for $T=0~{\rm MeV}$, Equation~\eqref{eq_s3:fe} leads to the equality between free and internal energy.
	
	The entropy density $s$, which appears in Equation~\eqref{eq_s3:fe}, has the same functional form as a noninteracting gas system, given by the equation
	\begin{align}
		s_{\tau}(n,T,I)=&-g\int \frac{d^3k}{(2\pi)^3}\left[f_{\tau} \ln
		f_{\tau}\right. \nonumber \\ &\left.+(1-f_{\tau}) \ln(1-f_{\tau})\right],
		\label{eq_s3:s}
	\end{align}
	where the spin degeneracy $g$ for protons, neutrons, electrons, and muons is equal to 2 and that for neutrinos is equal to 1.
	For the described thermodynamic system, pressure and chemical potentials are defined as follows:
	\begin{align}
		P&=-\frac{\partial E_{\rm tot}}{\partial V}\Bigg\vert_{S,N_i}=n^2\frac{\partial \left(\mathcal{E}/n\right)}{\partial n}\Bigg\vert_{S,N_i}, \\
		\mu_{i}&=\frac{\partial E_{\rm tot}}{\partial N_i}\Bigg\vert_{S,V,N_{j\neq i}}=\frac{\partial
			\mathcal{E}}{\partial n_i}\Bigg\vert_{S,V,n_{j\neq i}}.
		\label{eq_s3:p_and_m}
	\end{align}
	
	\subsection{Bulk Thermodynamic Quantities}
	It what follows, we will focus on the presentation of bulk thermodynamic quantities and approximations related to the present study. As the key quantity is the free energy, the pressure and chemical potentials are connected with the derivative of the total free energy $F_{\rm tot}$ and defined as
	\begin{align}
		P&=-\frac{\partial F_{\rm tot}}{\partial V}\Bigg\vert_{T,N_i}=n^2\frac{\partial \left(f/n\right)}{\partial
			n}\Bigg\vert_{T,N_i}, \\
		\mu_{i}&=\frac{\partial F_{\rm tot}}{\partial N_i}\Bigg\vert_{T,V,N_{j\neq i}}=\frac{\partial f}{\partial n_i}\Bigg\vert_{T,V,n_{j\neq i}},
		\label{eq_s3:p_and_m_isothermal}
	\end{align}
	where $f$ denotes the free energy density. Even more, the pressure $P$ can also be calculated from Goodstein~\citeyearpar{Goodstein-85} and Fetter \& Walecka~\citeyearpar{Fetter-03},
	\begin{equation}
		P=Ts-\mathcal{E}+\sum_{i}\mu_in_i.
		\label{eq_s3:p_isothrmal}
	\end{equation}
	The calculation of the entropy per particle $S(n,T)$ is done by differentiating the free energy density $f$ with respect to the temperature,
	\begin{equation}
		S(n,T)=- \frac{\partial \left(f/n\right)}{\partial T}\Bigg\vert_{V,N_i}=-\frac{\partial F}{\partial T}\Bigg\vert_{n}.
		\label{eq_s3:S_isothermal}
	\end{equation}
	The comparison between Equations~\eqref{eq_s3:s} and~\eqref{eq_s3:S_isothermal} for the entropy provides a testing criterion of the approximation used in the present work.
	
	By applying Equation~\eqref{eq_s3:p_and_m_isothermal}, the chemical potentials take the form (for a proof, see Prakash~\citeyear{Prakash-94}, as well as Nicotra et al.~\citeyear{AA.451.1.2010}; Burgio et al.~\citeyear{Burgio2007})
	\begin{subequations}
		\begin{align}
			\mu_n&=F+u\frac{\partial F}{\partial u}\Bigg\vert_{Y_p,T}-Y_p\frac{\partial F}{\partial Y_p}\Bigg\vert_{n,T}, \\
			\mu_p&=\mu_n+\frac{\partial F}{\partial Y_p}\Bigg\vert_{n,T}, \\
			\hat{\mu}&=\mu_n-\mu_p=-\frac{\partial F}{\partial Y_p}\Bigg\vert_{n,T}.
			\label{eq_s3:m_isothermal}
		\end{align}
	\end{subequations}
	The free energy $F(n,T,I)$ and the internal energy $E(n,T,I)$ can be expressed by the following parabolic approximations (PAs; Nicotra et al.~\citeyear{AA.451.1.2010}; Burgio et al.~\citeyear{Burgio2007}; Xu et al.~\citeyear{XU2007348}; Moustakidis~\citeyear{PhysRevC.78.054323}; Moustakidis \& Panos~\citeyear{PhysRevC.79.045806}):
	\begin{subequations}
		\begin{align}
			F(n,T,I)&=F(n,T,I=0)+I^2F_{\rm sym}(n,T),
			\label{eq_s3:F_parabolic} \\
			E(n,T,I)&=E(n,T,I=0)+I^2E_{\rm sym}(n,T),
			\label{eq_s3:E_parabolic}
		\end{align}
	\end{subequations}
	where
	\begin{subequations}
		\begin{align}
			F_{\rm sym}(n,T)= F(n,T,I=1)-F(n,T,I=0),
			\label{eq_s3:F_sym_parabolic} \\
			E_{\rm sym}(n,T)= E(n,T,I=1)-E(n,T,I=0).
			\label{eq_s3:E_sym_parabolic}
		\end{align}
	\end{subequations}
	In order to apply the above approximation, validity checking of the parabolic law is mandatory. The validity of the PA, at least in the present model, was tested previously. It has been proved that the PA is well satisfied not only on the internal energy, but also on the free energy (Moustakidis~\citeyear{PhysRevC.78.054323}; Moustakidis \& Panos~\citeyear{PhysRevC.79.045806}). A similar statement about the validity of the PA is also found in Burgio et al.~\citeyearpar{Burgio2007}, Nicotra et al.~\citeyearpar{AA.451.1.2010}, Xu et al.~\citeyearpar{XU2007348}. However, in other similar studies (Tan et al.~\citeyear{PhysRevC.93.035806}), it was found that the validity of the PA suffers from uncertainties. We conjecture that the validity of the PA strongly depends on the specific character of each nuclear model.
	
	The key quantity of Equation~\eqref{eq_s3:m_isothermal} can be obtained by using Equation~\eqref{eq_s3:F_parabolic} as
	\begin{equation}
		\hat{\mu}=\mu_n-\mu_p=4(1-2Y_p)F_{\rm sym}(n,T).
		\label{eq_s3:m_hat}
	\end{equation}
	This equation is similar to that obtained for cold catalyzed nuclear matter by replacing $E_{\rm sym}(n)$ with $F_{\rm sym}(n,T)$.
	
	It is intuitive to assume, based mainly on Equations~\eqref{eq_s3:F_parabolic} and~\eqref{eq_s3:E_parabolic}, that the entropy must also exhibit a quadratic dependence on the asymmetry parameter $I$; that is, according to the parabolic law (Moustakidis~\citeyear{doi:10.1142/S0218271809015023}),
	\begin{equation}
		S(n,T,I)=S(n,T,I=0)+I^2S_{\rm sym}(n,T),
		\label{eq_s3:S_parabolic}
	\end{equation}
	where
	\begin{align}
		S_{\rm sym}(n,T)&=S(n,T,I=1)-S(n,T,I=0) \nonumber \\
		&=\frac{1}{T}(E_{\rm sym}(n,T)-F_{\rm sym}(n,T)).
		\label{eq_s3:S_sym_parabolic}
	\end{align}
	
	\subsubsection{Lepton Contribution to EOS}
	In principle, the hot nuclear matter is composed, except for the two baryons (protons and neutrons), by photons and leptons (electrons, muons, and neutrinos) and their corresponding antiparticles (positrons, antimuons, and antineutrinos).
	
	In order to be stable, nuclear matter at high densities must be in chemical equilibrium for all reactions (including the weak interactions). Electron capture and $\beta$ decay would take place simultaneously as
	\begin{equation}
		p +e^{-}\longrightarrow n+ \nu_e \quad \text{and} \quad n \longrightarrow p+e^{-}+\bar{\nu}_e.
		\label{eq_s3:beta_decay}
	\end{equation}
	Both of them directly affect the EOS, as they change the electron per nucleon fraction $Y_{e}$. By assuming that the generated neutrinos have already left the system, the absence of neutrino trapping has a dramatic effect on the EOS, as a significant change in the values of the proton fraction $Y_{p}$ is in order (Takatsuka et al.~\citeyear{10.1143/ptp/92.4.779}; Takatsuka~\citeyear{10.1143/PTP.95.901}). The absence of neutrinos implies that
	\begin{equation}
		\hat{\mu}=\mu_n-\mu_p=\mu_e.
		\label{eq_s3:cp}
	\end{equation}
	In general, we consider that nuclear matter contains neutrons, protons, electrons, and muons. Muons decay to electrons as (Suh \& Mathews~\citeyear{Suh_2001})
	\begin{equation}
		\mu^{-} \longrightarrow e^{-} + \nu_{\mu} + \bar{\nu}_{e},
	\end{equation}
	but when the Fermi energy of the electrons approaches the muon rest mass $m_{\mu}\simeq 105.7~{\rm MeV}$ (due to their rest mass, it is expected to merely appear at the saturation nuclear density), it becomes energetically favorable for electrons at the top level of the Fermi sea to decay into muons with neutrinos and antineutrinos escaping from the star. Hence, above some density, muons and electrons are in an equilibrium state,
	\begin{equation}
		\mu^{-} \leftrightarrow e^{-},
	\end{equation}
	assuming that the neutrinos left the star. These particles are considered to be in a $\beta$-equilibrium state, where the following relations hold:
	\begin{equation}
		\mu_n=\mu_p+\mu_e, \quad \text{and} \quad \mu_e=\mu_{\mu}.
		\label{eq_s3:cps}
	\end{equation}
	The neutrality charge condition is also satisfied through the relation
	\begin{equation}
		n_p=n_e+n_{\mu}.
		\label{eq_s3:charge}
	\end{equation}
	The density of leptons (electrons and muons) is expressed through the relation
	\begin{equation}
		n_l=\frac{2}{(2\pi)^3}\int \frac{d^{3}k}{1+\exp\left[\frac{\sqrt{\hbar^2k^2c^2+m_l^2c^4}-\mu_l}{T}\right]}.
		\label{eq_s3:n_leptons}
	\end{equation}
	Equations~\eqref{eq_s3:m_hat}, and~\eqref{eq_s3:cps} -~\eqref{eq_s3:n_leptons} are solved in a self-consistent way for the calculation of the proton fraction $Y_p$, the lepton fractions $Y_e$ and $Y_{\mu}$, and the lepton chemical potentials $\mu_e$ and $\mu_{\mu}$ as functions of the baryon density $n$ for various values of the temperature $T$.
	
	Afterward, the energy density and pressure of leptons are calculated through the following formulae:
	\begin{equation}
		\mathcal{E}_{l}(n_l,T)=\frac{2}{(2\pi)^3}\int \frac{d^{3}k~\sqrt{\hbar^2 k^2 c^2+m_l^2c^4}}{1+\exp\left[\frac{\sqrt{\hbar^2k^2c^2+m_l^2c^4}-\mu_l}{T}\right]},
		\label{eq_s3:e_leptons}
	\end{equation}
	\begin{align}
		P_l(n_l,T)&=\frac{1}{3}\frac{2(\hbar c)^2}{(2\pi)^3}\int \frac{1}{\sqrt{\hbar^2 k^2 c^2+m_l^2c^4}} \nonumber \\
		&\times \frac{d^{3}k~k^2}{1+\exp\left[\frac{\sqrt{\hbar^2k^2c^2+m_l^2c^4}-\mu_l}{T}\right]}.
		\label{eq_s3:p_leptons}
	\end{align}
	
	According to Equations~\eqref{eq_s3:m_hat} and~\eqref{eq_s3:cps}, the chemical potentials of electrons and muons, which are equal, are
	\begin{equation}
		\mu_{e}=\mu_{\mu}=\mu_n-\mu_p=4I(n,T)F_{\rm sym}(n,T).
		\label{eq_s3:cp_leptons}
	\end{equation}
	Equation~\eqref{eq_s3:cp_leptons} is crucial for the calculation of the proton fraction as a function of the baryon density and for various temperatures. The EOS of hot nuclear matter in the $\beta$-equilibrium state is provided through the calculation of the total energy density $\mathcal{E}_{\rm t}$, as well as the total pressure $P_{\rm t}$. The total energy density is given by
	\begin{align}
		\mathcal{E}_{\rm t}(n,T,I)=&\mathcal{E}_b(n,T,I)+\sum_{l}\mathcal{E}_l(n,T,I) \nonumber \\&+\sum_{\bar{l}}\mathcal{E}_{\bar{l}}(n,T,I)+\mathcal{E}_{\gamma}(n,T),
		\label{eq_s3:e_total}
	\end{align}
	where $\mathcal{E}_b(n,T,I)$, $\mathcal{E}_l(n,T,I)$, $\mathcal{E}_{\bar{l}}(n,T,I)$, and $\mathcal{E}_{\gamma}(n,T)$ are the contributions of baryons, particles and antiparticles of leptons, and photons, respectively. The total pressure is
	\begin{align}
		P_{\rm t}(n,T,I)=&P_b(n,T,I)+\sum_{l}P_l(n,T,I) \nonumber \\&+\sum_{\bar{l}}P_{\bar{l}}(n,T,I)+P_{\gamma}(T),
		\label{eq_s3:p_total}
	\end{align}
	where $P_b(n,T,I)$ is the contribution of baryons (see Equation~\eqref{eq_s3:p_isothrmal}),
	\begin{align}
		P_b(n,T,I)= &T\sum_{\tau=p,n}s_{\tau}(n,T,I) \nonumber \\ &+\sum_{\tau=n,p}n_{\tau}\mu_{\tau}(n,T,I)-\mathcal{E}_b(n,T,I),
		\label{eq_s3:p_baryon}
	\end{align}
	while $P_l(n,T,I)$, $P_{\bar{l}}(n,T,I)$, and $P_{\gamma}(T)$ are the contributions of particles and antiparticles of leptons and photons, respectively.
	
	It is worth mentioning that, in principle, it is necessary to include photons and antiparticles, which are in thermal equilibrium with the other constituents of the hot nuclear matter. However, in the present study, we excluded them, since their contribution is negligible (Takatsuka et al.~\citeyear{10.1143/ptp/92.4.779}).
	
	\subsection{Isothermal Temperature Profile}
	In the present study, we take under consideration that nuclear matter consists only of neutrons, protons, and electrons. Therefore, electrons are the only leptons that contribute to the energy density and pressure. Assuming that, for each value of temperature, the proton fraction is a well-known function of the baryon density, $Y_{p}=Y_{p}(n)$, the total energy density reads as
	\begin{equation}
		\mathcal{E}_{\rm t}(n,T,Y_{p})=\mathcal{E}_b(n,T,Y_{p})+\mathcal{E}_e(n,T,Y_{p}),
		\label{eq_s3:e_total_1}
	\end{equation}
	where
	\begin{equation}
		\mathcal{E}_b(n,T,Y_{p})=nF_{\rm PA} + nTS_{\rm PA},
		\label{eq_s3:e_baryon_1}
	\end{equation}
	$\mathcal{E}_e(n,T,Y_{p})$ is given by Equation~\eqref{eq_s3:e_leptons} replacing the leptons with electrons and  $\mu_{e}$ from Equation~\eqref{eq_s3:cp_leptons}, and, in the frame of the PA, $F_{\rm PA}$ and $S_{\rm PA}$ are given by Equations~\eqref{eq_s3:F_parabolic} and~\eqref{eq_s3:S_parabolic}, respectively.
	In addition, the total pressure reads as
	\begin{equation}
		P_{\rm t}(n,T,Y_{p})=P_b(n,T,Y_{p})+P_e(n,T,Y_{p}),
		\label{eq_s3:p_total_1}
	\end{equation}
	where
	\begin{equation}
		P_b(n,T,Y_{p})=n^2\frac{\partial F_{\rm PA}(n,T,Y_{p})}{\partial n}\Bigg\vert_{T,n_i},
		\label{eq_s3:p_baryon_1}
	\end{equation}
	and  $P_e(n,T,Y_{p})$ is given by Equation~\eqref{eq_s3:e_leptons} replacing the leptons with the electrons and $\mu_{e}$ from Equation~\eqref{eq_s3:cp_leptons}.
	
	Henceforth, in the present study, Equation~\eqref{eq_s3:e_total_1} for the energy density and Equation~\eqref{eq_s3:p_total_1} for the pressure are the ingredients for the construction of isothermal EOSs of hot nuclear matter in a $\beta$-equilibrium state.
	
	\subsubsection{Thermal Index}
	Except protoneutron stars and supernovae, hot EOSs find their place in neutron star mergers where the increase of temperature is rather significant. A usual treatment, in order to study the effects of temperature on neutron stars and to include thermal effects in neutron star merger simulations, is the effective thermal index, defined as (Constantinou et al.~\citeyear{PhysRevC.89.065802}, ~\citeyear{PhysRevC.92.025801})
	\begin{equation}
		\Gamma_{\rm th}(n) = 1 + \frac{P_{\rm th}(n)}{\mathcal{E}_{\rm th}(n)},
		\label{eq_s9_3:gamma_thermal}
	\end{equation}
	where $P_{\rm th}(n)$ and $\mathcal{E}_{\rm th}(n)$ are the pressure and energy density contribution to the cold EOS due to temperature. More precisely, for a specific value of temperature, the right-hand side terms of Equation~\eqref{eq_s9_3:gamma_thermal} are defined as
	\begin{subequations}
		\begin{align}
			P_{\rm th}(n) &= P(T,n) - P(T=0,n), \\
			\mathcal{E}_{\rm th}(n) &= \mathcal{E}(T,n) - \mathcal{E}(T=0,n)
			\label{eq_s9_3:pressure_energy_thermal}.
		\end{align}
	\end{subequations}
	
	It has to be noted that although Equation~\eqref{eq_s9_3:gamma_thermal} is artificially and not self-consistently constructed, it has been widely used in order to introduce the effects of temperature in isothermal EOSs (Bauswein et al.~\citeyear{PhysRevD.82.084043}).
	
	In most cases, the values of the thermal index are taken to be constant, an approximation that seems to be unrealistic, since a high density dependence is suggested by the interactions of cold catalyzed matter.
	
	\subsection{Isentropic Temperature Profile and Neutrino Trapping}
	In the case of the isentropic profile, we consider that the entropy per baryon and lepton fraction are fixed in the interior of a protoneutron star. In particular, according to Equation~\eqref{eq_s3:beta_decay}, we consider that neutrinos are trapped in the interior of the star, a process that leads to a dramatic increase of the proton fraction. Now the chemical equilibrium can be expressed in terms of the chemical potentials for the four species,
	\begin{equation}
		\mu_n+\mu_{\nu_e}=\mu_p+\mu_e.
	\label{chem-equil-PN-1}
	\end{equation}  
	Obviously, the charge neutrality demands $Y_p=Y_e$, while the total fraction of leptons reads as $Y_l=Y_e+Y_{\nu_e}$. Moreover, the chemical equilibrium leads to the expression
	\begin{equation}
		\mu_e-\mu_{\nu_e}=\mu_n-\mu_p=4(1-2Y_p)F_{\rm sym}(n,T).
		\label{chem-equil-PN-2}
	\end{equation}
	Similar to the isothermal profile, one can self-consistently solve the relevant equations in order to calculate the density and temperature dependence of proton and neutrino fractions, as well as the corresponding chemical potentials for a fixed value of the total entropy per baryon. However, in order to avoid computational complications (arising mainly from the system of the coupled integral equations), we follow the approximation introduced by Takatsuka et al.~\citeyearpar{10.1143/ptp/92.4.779}. In particular, it was found that the proton fraction is well approximated (within $3 \%$ accuracy) by the empirical formula $Y_p\simeq 2/3 Y_l+0.05$.	The ingredients for the construction of isentropic EOSs are given by Equations~\eqref{eq_s3:e_total} and ~\eqref{eq_s3:p_total}.
		
	Two important quantities related to the measure of stiffness of the EOS and, consequently, the stability of protoneutron stars are the adiabatic index, defined as
	\begin{equation}
		\Gamma = \frac{n}{P} \frac{\partial P}{\partial n}\Bigg\vert_{S},
		\label{adiabatic index}
	\end{equation}
	and the speed of sound given by Landau \& Lifshitz~\citeyearpar{1969stph.book.....L}
	\begin{equation}
		\frac{c_{s}}{c} = \sqrt{\frac{\partial P}{\partial \mathcal{E}}}\Bigg\vert_{S}.
	\end{equation}
	
	\section{Rapidly Rotating Hot Neutron Stars} \label{section:Rapidly rotating hot neutron stars}
	Einstein's equations for a rigidly rotating neutron star are the most suitable tool to describe its macroscopic properties. In this case, the metric for curved spacetime is (Weber~\citeyear{Weber-1996}; Glendenning~\citeyear{Glendenning-2000})
	\begin{align}
		ds^{2} = &-e^{2\nu}dt^{2} + e^{2\phi}\left(d\varphi - N^{\varphi}dt\right)^{2} \nonumber \\
		&+ e^{2\omega}\left(dr^{2} + r^{2}d\theta^{2}\right),
		\label{eq_s4:metric}
	\end{align}
	where $\nu$, $\phi$, $N^{\varphi}$, and $\omega$ are metric functions that depend on the coordinates $r$ and $\theta$. These equations are solved numerically, coupled to the hydrostatic equilibrium condition, and with source terms given by that of a perfect fluid. The latter is possible if we neglect sources of nonisotropic stresses, as well as viscous ones, and heat transport. The energy-momentum tensor that describes the perfect fluid is
	\begin{equation}
		T^{\mu\nu} = \left(\mathcal{E} + P\right)u^{\mu}u^{\nu} + Pg^{\mu\nu},
		\label{eq_s4:fluid}
	\end{equation}
	where $u^{\mu}$ and $u^{\nu}$ are the fluid's four-velocity. The thermodynamical quantities, energy density and pressure, are denoted as $\mathcal{E}$ and $P$, respectively, and $g^{\mu\nu}$ denotes the spacetime  metric function.
	
	The stability of cold rotating neutron stars is acquired via the turning-point criterion, which is only a sufficient and not a necessary one. In fact, the neutral stability line is positioned to the left of the turning-point line in $\left(M,\rho_{c}\right)$ space. The latter indicates that the star will collapse before reaching the turning-point line (Takami et al.~\citeyear{10.1111/j.1745-3933.2011.01085.x}; Weih et al.~\citeyear{10.1093/mnrasl/slx178}).
	
	\subsection{Instabilities in Hot Neutron Stars}
	\begin{figure*}
		\includegraphics[width=\textwidth]{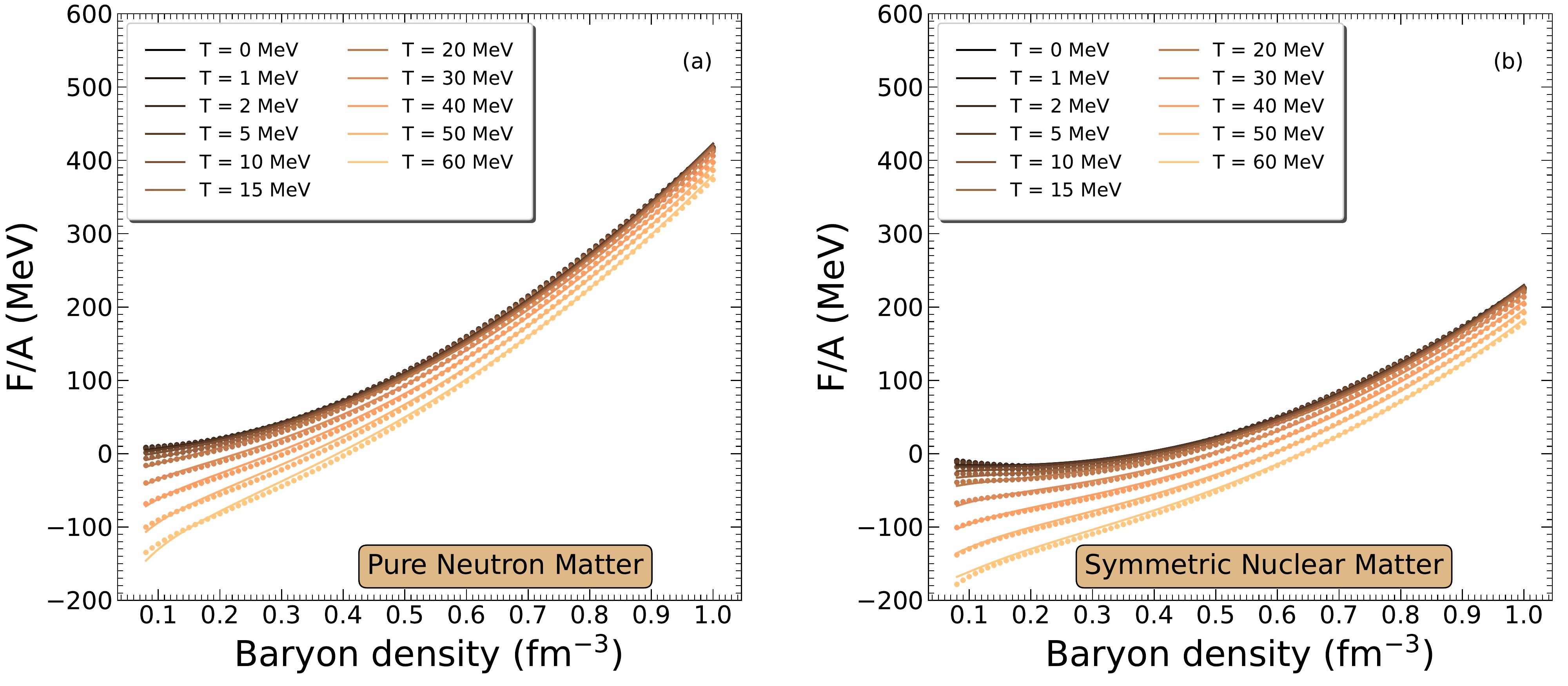}
		\caption{Free energy per particle as a function of baryon density for (a) PNM and (b) SNM for temperatures in the range $[0,60]$ MeV and the MDI+APR1 EOS. Data and fits are presented by circles and solid lines, respectively. (The following figures also refer to the MDI+APR1 EOS.)}
		\label{fig:free energy}
	\end{figure*}

	The stability of hot neutron stars is acquired via a specific version of the secular instability criterion of Friedman et al.~\citeyearpar{ApJ.325.722}, which follows Theorem I of Sorkin~\citeyearpar{ApJ.257.847}. We choose a continuous sequence of equilibria to be at a fixed baryon number $N_{\rm bar}$ and total entropy of the neutron star $S_{\rm t}^{\rm ns}$, and the extremal point of the stability loss is (Goussard et al.~\citeyear{ApJ.321.822})
	\begin{equation}
		\frac{\partial J}{\partial n_{b}^{c}}\Bigg\vert_{N_{\rm bar},S_{\rm t}^{\rm ns}} = 0,
	\end{equation}
	where $J$ and $n_{b}^{c}$ are the angular momentum and central baryon density of the star, respectively.
	
	In addition, a turning point in the sequence occurs where three out of four derivatives, $\partial M_{\rm gr}/\partial n_{b}^{c}$, $\partial M_{\rm b}/\partial n_{b}^{c}$, $\partial J/\partial n_{b}^{c}$, and $\partial S_{\rm t}^{\rm ns}/\partial n_{b}^{c}$, where $M_{\rm gr}$ and $M_{\rm b}$ denote the gravitational and baryon mass, vanish (Kaplan et al.~\citeyear{Kaplan_2014}; Marques et al.~\citeyear{PhysRevC.96.045806}). At this point, the turning-point theorem shows that the fourth derivative also vanishes, and the sequence has transitioned from stable to unstable.
	
	The criterion for distinguishing secularly stable from unstable configurations is meaningful only for constant entropy per baryon or temperature (Marques et al.~\citeyear{PhysRevC.96.045806}). In our calculations, as the entropy per baryon and temperature are constant throughout the star, the other three criteria simultaneously vanish at the maximum mass configuration, which is the last stable point. It has to be mentioned that the rotating configuration with maximum mass and the one with maximum angular velocity do not generally coincide (Friedman \& Stergioulas~\citeyear{friedman_stergioulas_2013}). However, the difference is very small, and it could not be detected within the precision of our calculations (Goussard et al.~\citeyear{ApJ.321.822}).
	
	For the numerical integration of the equilibrium equations, we used the publicly available numerical code \emph{nrotstar} from the C++ Lorene/Nrotstar library (LORENE~\citeyear{lorene}) (for more details, see Section~\ref{section: numerical code}).
	
	\section{Discussion and Conclusions} \label{section:Discussion and Conclusions}
	\subsection{Free Energy and Proton Fraction}
	A key quantity related to the calculation of the proton fraction via $\beta$-equilibrium is the free energy per particle. Figure~\ref{fig:free energy} displays the free energy per particle as a function of the baryon density for temperatures in the range $[0,60]~{\rm MeV}$ and the MDI+APR1 EOS for both (a) PNM and (b) SNM (in the following, we refer only to the MDI+APR1 EOS). As is expected due to the quantum character of the hadronic matter, thermal effects are more pronounced at low densities, while at high densities, there is a tendency for convergence. Moreover for practical reasons, it is convenient to have analytical expressions for the dependence of the free energy on both baryon density and temperature. Following the suggestion of Lu et al.~\citeyearpar{PhysRevC.100.054335}, we employed the following functional form,
	\begin{align}
		\frac{F}{A}\left(n,T\right) = a_{0} &+ \left(a_{1} + a_{2}t^{2}\right) n + a_{3}n^{a_{4}} \nonumber \\ &+ a_{5}t^{2}ln(n) + \left(a_{6}t^{2} + a_{7}t^{a_{8}}\right)/n,
		\label{eq:fits_pnm_snm}
	\end{align}
	where $t=T/100~{\rm MeV}$, and $F/A$ and $n$ are given in units of $\rm MeV$ and $\rm fm^{-3}$, respectively. The parameters $a_i$ of the fit, with $i=0-8$, for the SNM and PNM are listed in Table~\ref{tab:parameters_of_pnm_snm}.
	
	\begin{figure}
		\includegraphics[width=\columnwidth]{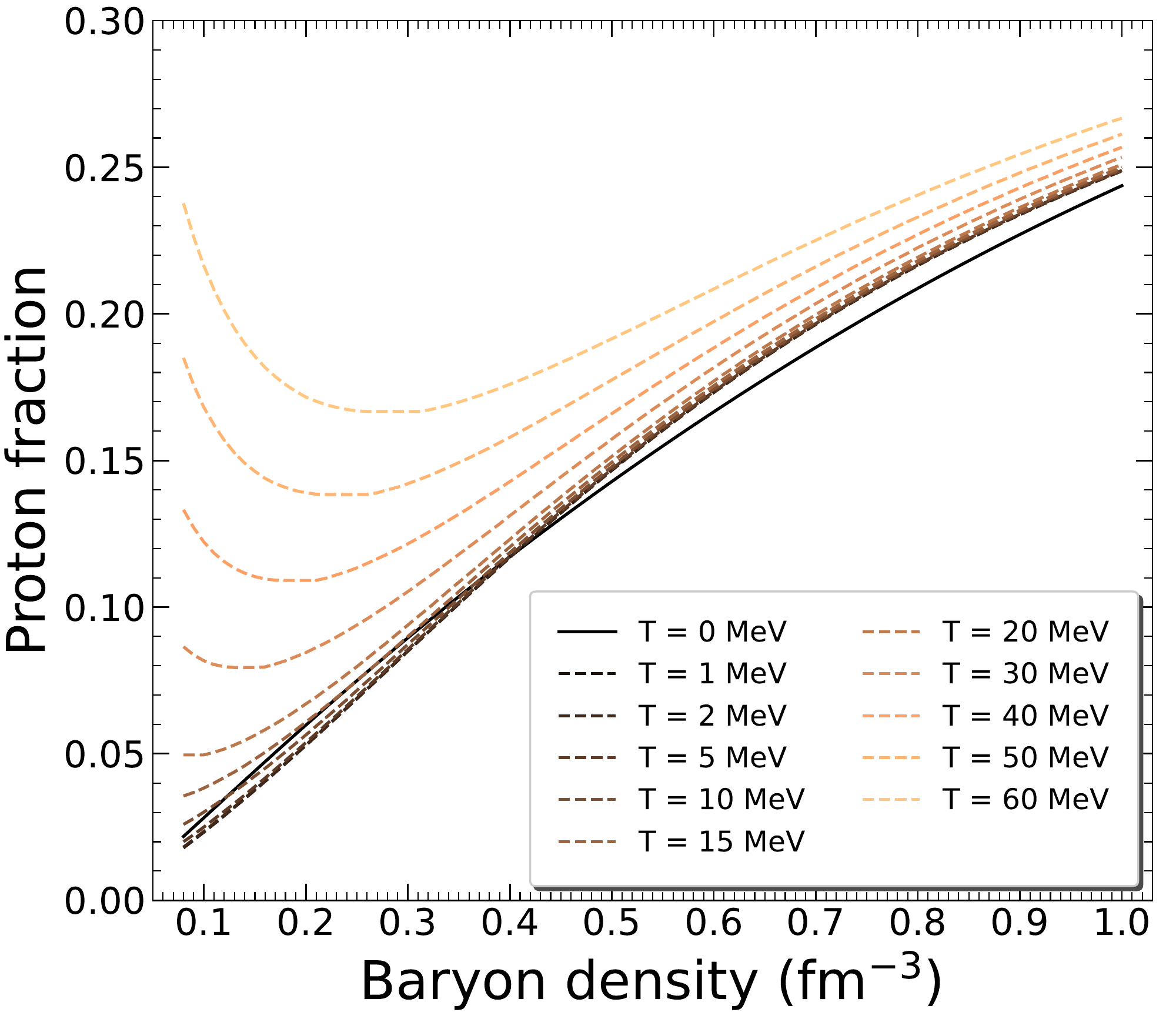}
		\caption{Proton fraction as a function of baryon density for temperatures in the range $[0,60]$ MeV. The cold configuration is presented by the solid line, while hot configurations are presented by the dashed ones.}
		\label{fig:proton fraction}
	\end{figure}

	\begin{table}
		\footnotesize
		\caption{Parameters of Equation~\eqref{eq:fits_pnm_snm} for PNM and SNM of MDI+APR1 EOS}
		\begin{tabular*}{\columnwidth}{@{\extracolsep{\fill} }  l  r  r }
			\toprule
			\toprule
			Parameters & PNM & SNM \\
			\midrule
			\decimals
			$a_{0}$ & 0.000 & -12.000 \\
			$a_{1}$ & 37.814 & -54.000 \\
			$a_{2}$ & -117.379 & -140.000 \\
			$a_{3}$ & 385.000 & 296.000 \\
			$a_{4}$ & 2.079 & 2.261 \\
			$a_{5}$ & 150.000 & 211.000 \\
			$a_{6}$ & -90.000 & -64.000 \\
			$a_{7}$ & 94.000 & 88.000 \\
			$a_{8}$ & 2.140 & 2.350 \\
			\bottomrule
		\end{tabular*}
		\label{tab:parameters_of_pnm_snm}
	\end{table}
	
	\begin{figure}
		\includegraphics[width=\columnwidth]{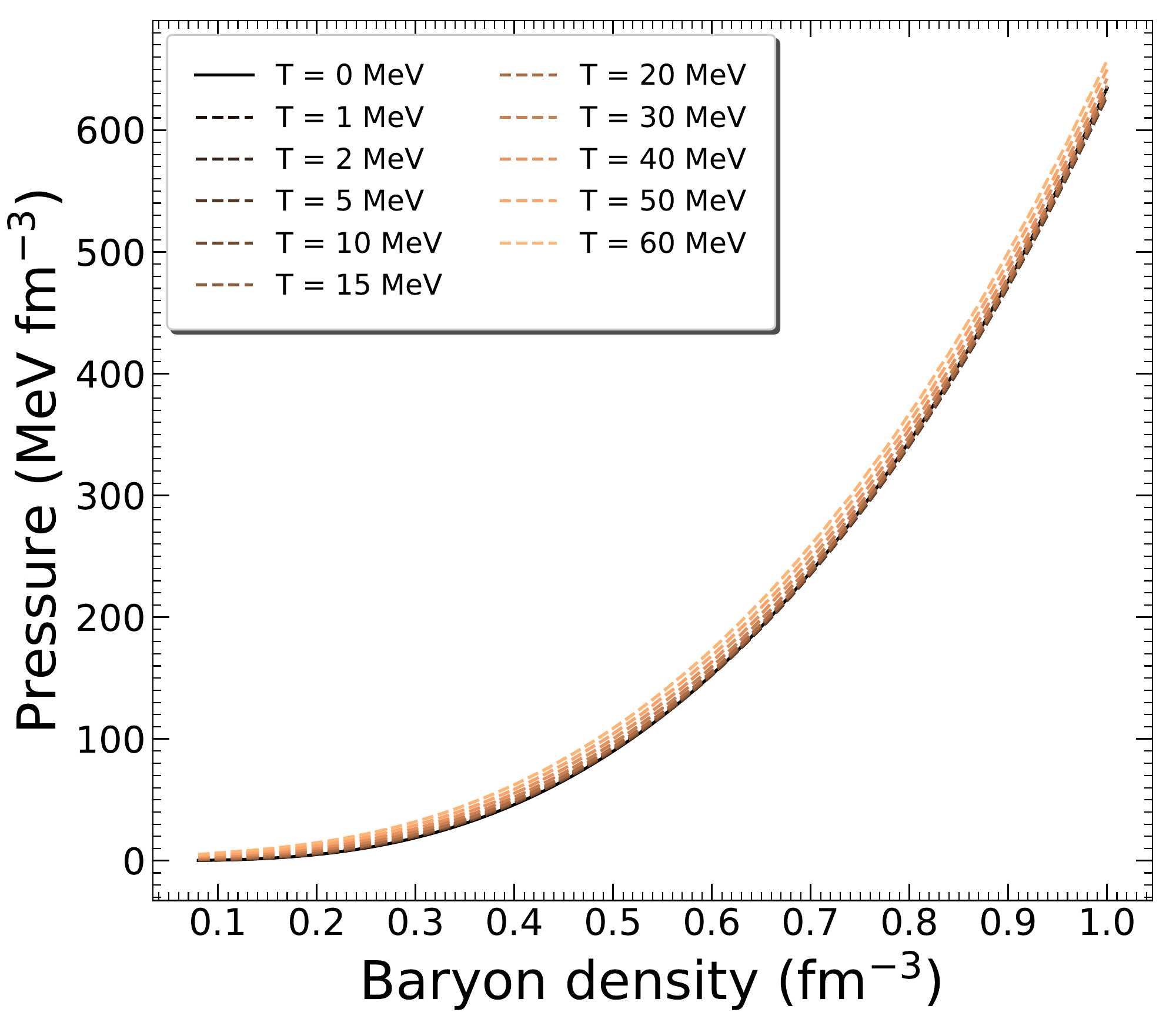}
		\caption{Pressure as a function of baryon density for temperatures in the range $[0,60]$ MeV. The cold configuration is presented by the solid line, while hot configurations are presented by the dashed ones.}
		\label{fig:pressure vs baryon density}
	\end{figure}
	
	Equation~\eqref{eq:fits_pnm_snm} is an excellent parameterization of the free energy per particle in the range of density $0.08~{\rm fm^{-3}}\leq n \leq 1~{\rm fm^{-3}}$ and temperature $0~{\rm MeV}\leq T \leq 60~{\rm MeV}$. In addition, through Equation~\eqref{eq_s3:S_isothermal}, we confirmed the very good accuracy between the analytical and numerical calculation of the entropy from Equation~\eqref{eq_s3:s}.
	
	The knowledge of the proton fraction is very important, since it is related not only to the specific structure of a neutron star but also to the direct (nucleonic) URCA process (Yakovlev \& Pethick~\citeyear{doi:10.1146/annurev.astro.42.053102.134013}). Figure~\ref{fig:proton fraction} displays the proton fraction as a function of the baryon density for temperatures in the range $[0,60]~{\rm MeV}$. Our predictions are very close to those found recently in Lu et al.~\citeyearpar{PhysRevC.100.054335}, where the authors employed a different nuclear model and approach. In particular, while in the low-density region, the proton fraction is very sensitive to the temperature, in the high-density region, the thermal effects are very mild. This is a direct consequence of the similar sensitivity of the free energy per particle to temperature shown in Figure~\ref{fig:free energy}. Furthermore, at a high temperature, the free symmetry energy plays an insignificant role, and, consequently, the nuclear system tends to become more symmetric.
	
	\subsection{EOS and Thermal and Adiabatic Indices}

	\begin{figure}
		\includegraphics[width=\columnwidth]{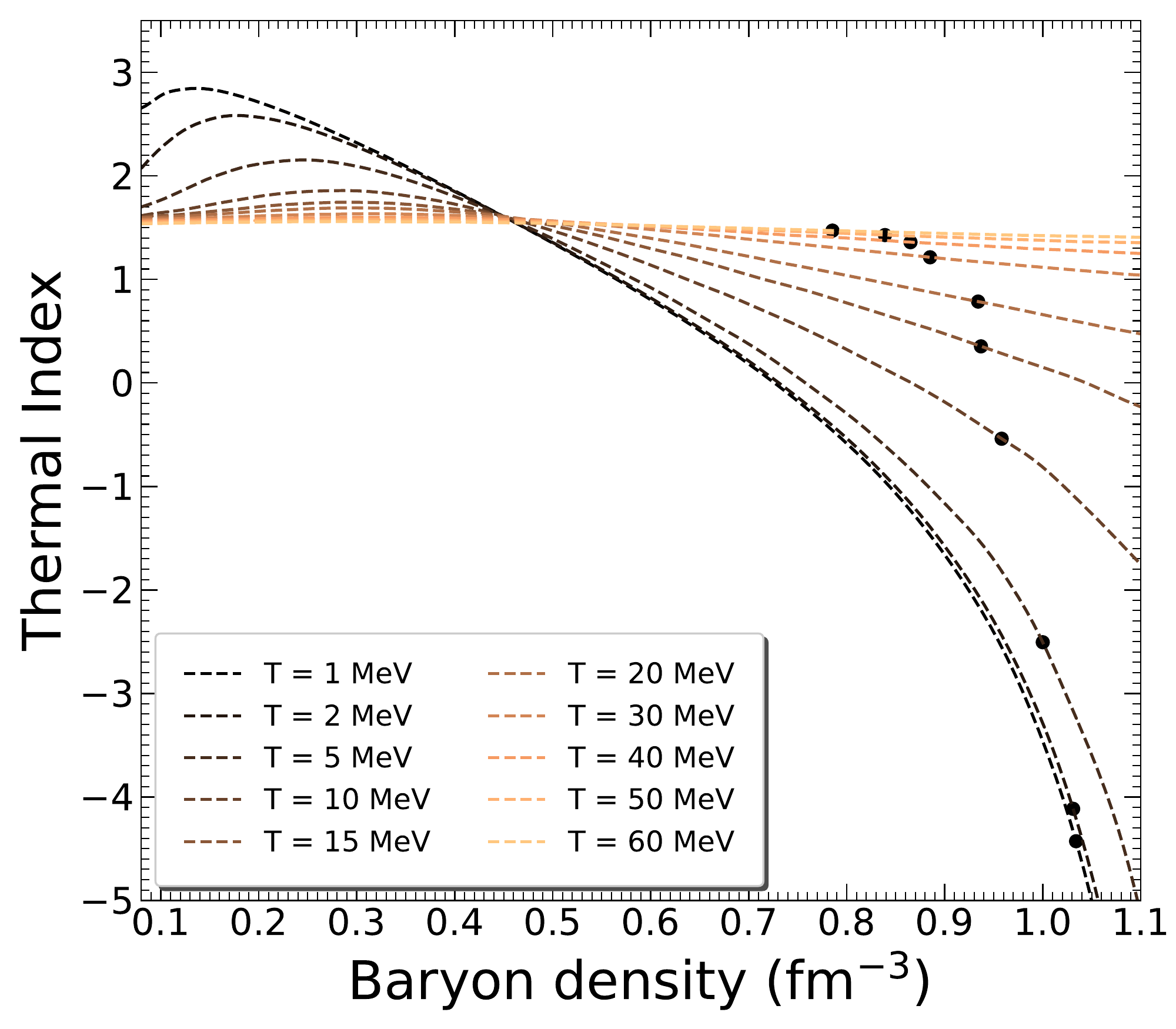}
		\caption{Thermal index as a function of baryon density for temperatures in the range $[1,60]$ MeV. Black circles represent the central baryon density at which the maximum mass configuration appears.}
		\label{fig:thermal index}
	\end{figure}
	
	Figure~\ref{fig:pressure vs baryon density} displays the pressure as a function of the baryon density for temperatures in the range $[0,60]~{\rm MeV}$. In particular, we present one EOS for the cold catalyzed matter and 10 isothermal ones.
	
	Furthermore, we study the thermal index, a quantity that fully relies on the energy and pressure thermal components. In Figure~\ref{fig:thermal index}, we display the thermal index as a function of the baryon density for temperatures in the range $[1,60]~{\rm MeV}$. An important density dependence is clearly presented, especially for temperatures in the range $[1,30]~{\rm MeV}$. At higher temperatures ($T > 30~{\rm MeV}$), the thermal index has an almost constant value, as its density dependence is rather insignificant.
	
	\begin{figure}
		\includegraphics[width=\columnwidth]{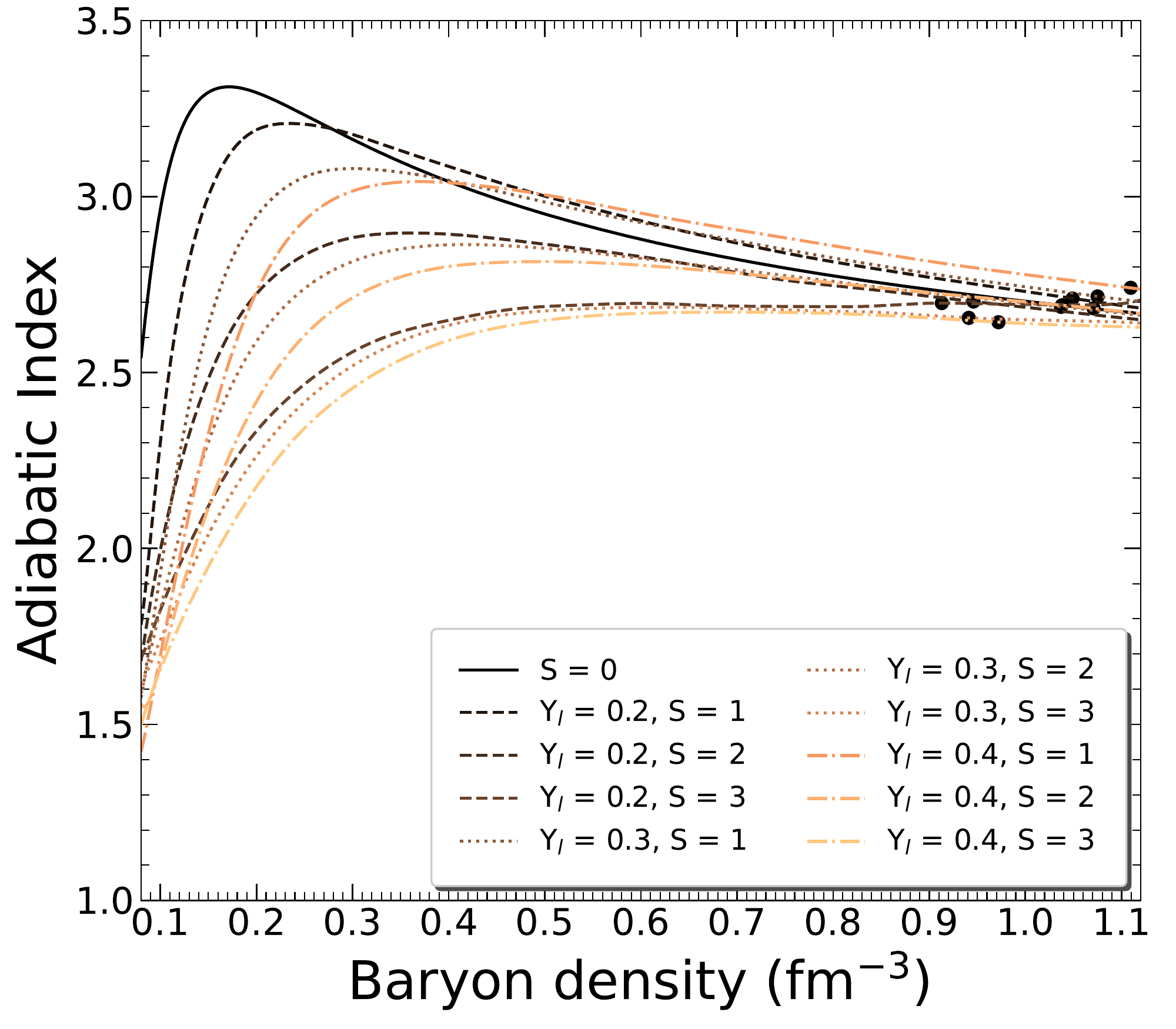}
		\caption{Adiabatic index as a function of baryon density for lepton fractions and entropies per baryon in the ranges $[0.2,0.4]$ and $[1,3]~k_{B}$, respectively. Black circles represent the central baryon density at which the maximum mass configuration appears. The cold configuration is presented by the solid line.}
		\label{fig:adiabatic index}
	\end{figure}
	
	We note here that due to the thermal effects that we analyzed, Equation~\eqref{eq_s9_3:gamma_thermal} might be strongly violated, in particular for EOSs with low values of temperatures ($T \leq 10~{\rm MeV}$) and, as a consequence, low values of proton fraction, where the energy density and pressure thermal components might even become negative (Lu et al.~\citeyear{PhysRevC.100.054335}).
	
	In the case of isentropic EOSs, we study both the adiabatic index and the speed of sound. In Figure~\ref{fig:adiabatic index}, we display the adiabatic index as a function of the baryon density for lepton fractions and entropies per baryon in the ranges $[0.2,0.4]$ and $[1,3]~k_{B}$, respectively. For a constant lepton fraction, the decreasing of the entropy per baryon leads to higher values of the central baryon density at which the maximum mass appears.
		
\begin{figure}
		\includegraphics[width=\columnwidth]{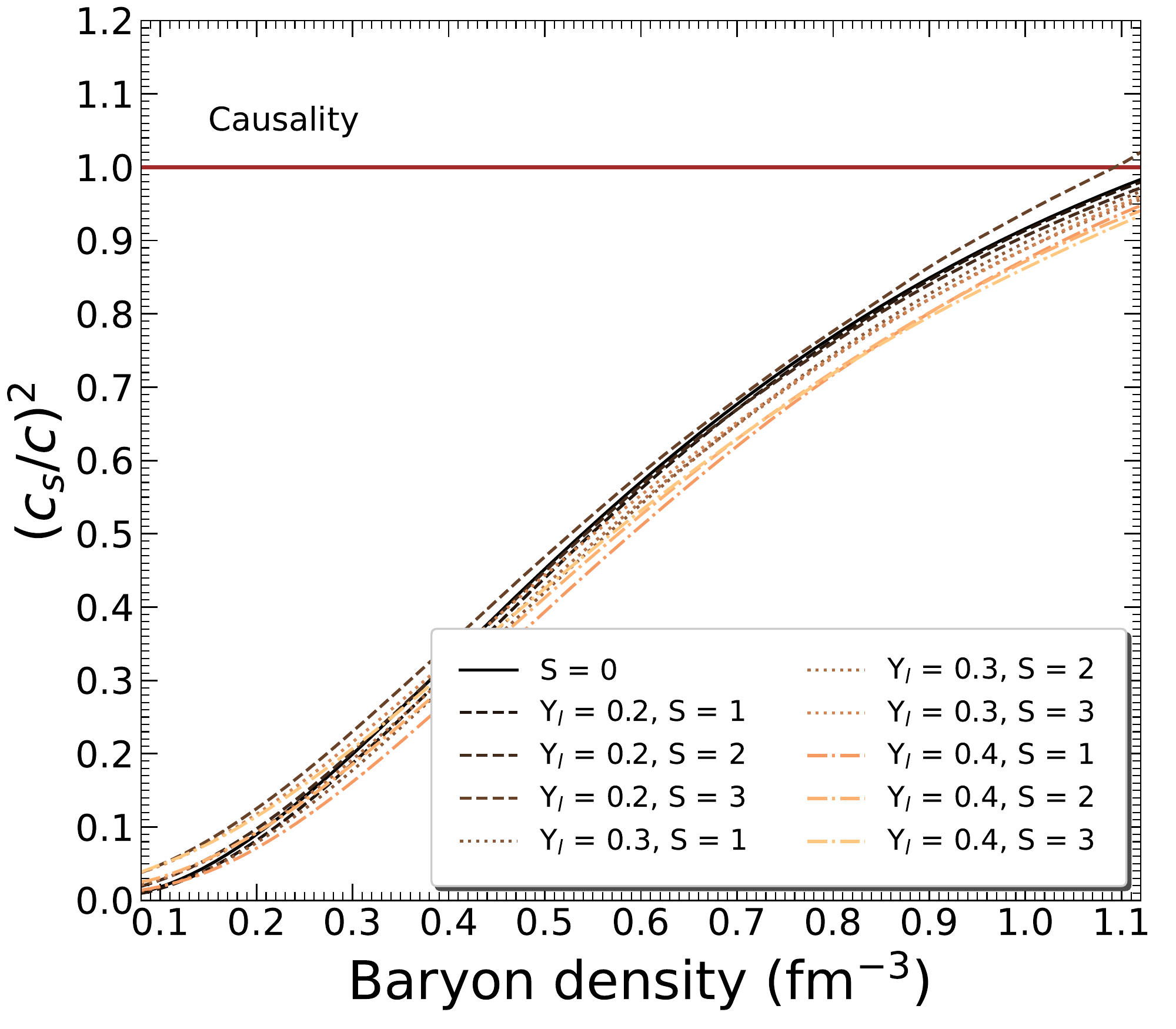}
		\caption{Square speed of sound in units of speed of light as a function of baryon density for lepton fractions and entropies per baryon in the ranges $[0.2,0.4]$ and $[1,3]~k_{B}$, respectively. The cold configuration is presented by the solid line.}
		\label{fig:speed of sound}
	\end{figure}

	In addition, in Figure~\ref{fig:speed of sound}, we present the square speed of sound in units of speed of light as a function of the baryon density. In this scenario, no EOSs, including the one with cold catalyzed matter, ever exceed the causality limit (see also Heiselberg \& Hjorth-Jensen~\citeyear{HEISELBERG2000237}). It has to be emphasized that one of the major advantages of the MDI model is to prevent the EOS from reaching the causality point. The latter is effective even to higher values of neutron star baryon density than the ones that correspond to the maximum mass configuration.
	
	\subsection{Thermal Effects on Nonrotating Neutron Stars}
	\begin{figure}
		\includegraphics[width=\columnwidth]{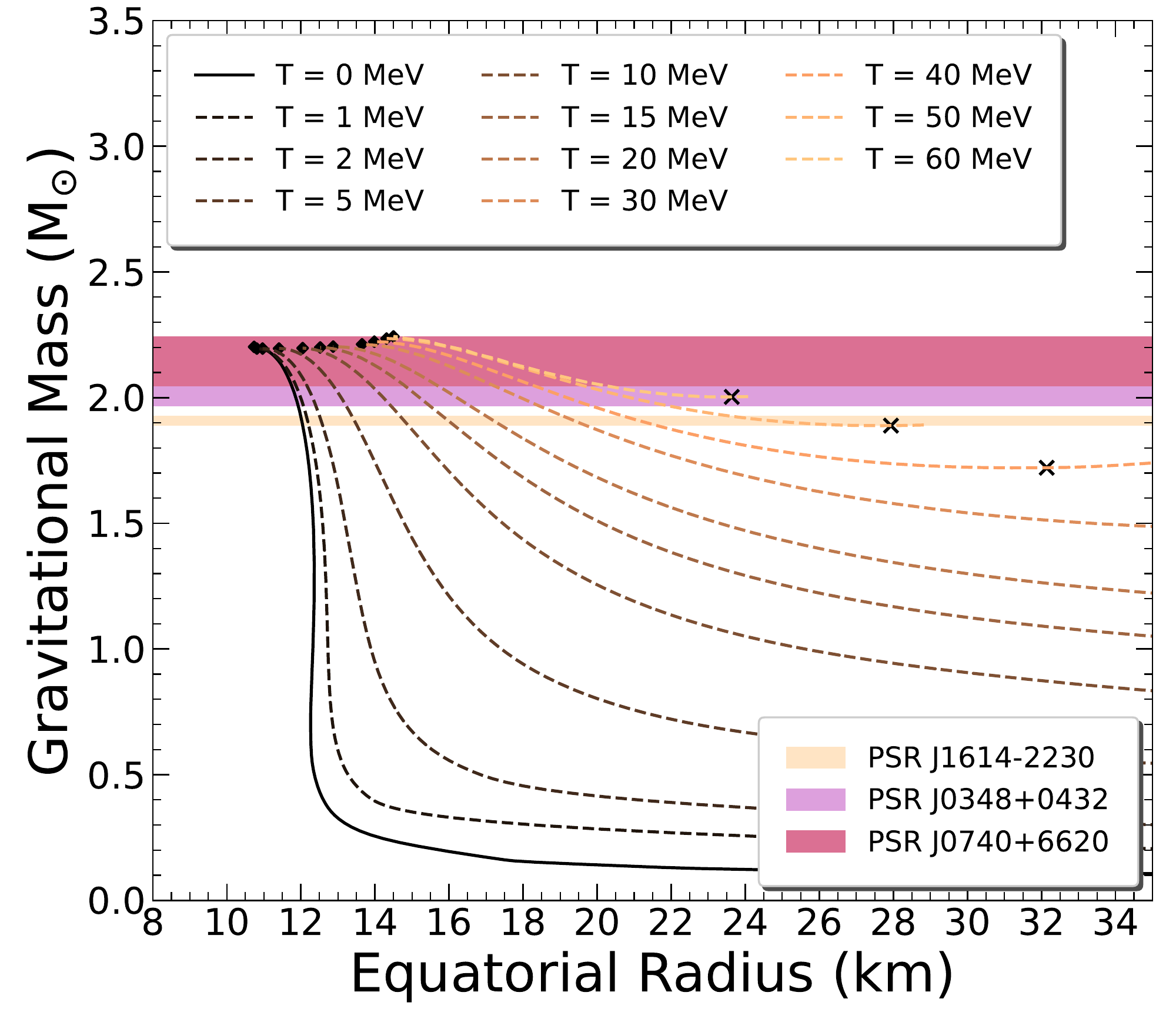}
		\caption{Gravitational mass as a function of equatorial radius for temperatures in the range $[0,60]$ MeV at the nonrotating configuration. The cold configuration is presented by the solid line, while hot configurations are presented by the dashed ones. The shaded regions from bottom to top represent the PSR J1614-2230 (Arzoumanian et al.~\citeyear{Arzoumanian-2018}), PSR J0348+0432 (Antoniadis et al.~\citeyear{Antoniadis-2013}), and PSR J0740+6620 (Cromartie et al.~\citeyear{Cromartie-2019}) pulsar observations for possible maximum mass. Black diamonds correspond to the maximum mass configuration in each case, while black crosses correspond to the minimum mass configuration. (The remaining minimum masses are positioned at higher values of equatorial radius.)}
		\label{fig:mass radius static}
	\end{figure}
	
	We now concentrate our study on the bulk properties of nonrotating neutron stars at the maximum mass configuration. In Figure~\ref{fig:mass radius static}, we display the gravitational mass as a function of the corresponding equatorial radius for temperatures in the range $[0,60]~{\rm MeV}$. It is worth clarifying that the nonhomogeneous nuclear matter phase disappears when the temperature is higher than $T\sim 15~{\rm MeV}$. To be more specific, the critical temperature $T_c$ where this transition (known as liquid-gas phase transition) is achieved is model-dependent. However, a well accepted value is close to $T_c=15~{\rm MeV}$ (Shen et al.~\citeyear{SHEN1998435}; Haensel et al.~\citeyear{Haensel-2007}).
	
	\begin{table}
		\footnotesize
		\caption{Summary of Nonrotating Isothermal Neutron Star Bulk Properties}
		\begin{tabular*}{\columnwidth}{@{\extracolsep{\fill} }  l  c  c  c  c  c }
			\toprule
			\toprule
			$T$ & $M_{b}^{\rm max}$ & $M_{gr}^{\rm max}$ & $R_{\rm max}$ & $n_{b}^{c}$ & $R_{1.4}$ \\
			(MeV) & $(M_{\odot})$ & $(M_{\odot})$ & (km) & ${\rm (fm^{-3})}$ & (km) \\
			\midrule
			\decimals
			0 & 2.622 & 2.202 & 10.734 & 1.038 & 12.353 \\
			1 & 2.599 & 2.195 & 10.809 & 1.034 & 12.633 \\
			2 & 2.567 & 2.195 & 10.963 & 1.031 & 13.321 \\
			5 & 2.501 & 2.195 & 11.407 & 1.000 & 15.150 \\
			10 & 2.427 & 2.197 & 12.044 & 0.958 & 18.315 \\
			15 & 2.380 & 2.199 & 12.520 & 0.937 & 21.676 \\
			20 & 2.345 & 2.203 & 12.869 & 0.934 & 25.922 \\
			30 & 2.310 & 2.212 & 13.650 & 0.885 & -  \\
			40 & 2.307 & 2.223 & 13.981 & 0.865 & - \\
			50 & 2.313 & 2.235 & 14.312 & 0.839 & - \\
			60 & 2.342 & 2.244 & 14.497 & 0.785 & - \\
			\bottomrule
		\end{tabular*}
		\label{tab:nonrotating_data}
		\tablecomments{\footnotesize Reported are the temperature $T$, baryon mass $M_{b}^{\rm max}$, gravitational mass $M_{\rm gr}^{\rm max}$, equatorial radius $R_{\rm max}$, and central baryon density $n_{b}^{c}$. The above properties correspond to the maximum gravitational mass configuration. The equatorial radius $R_{1.4}$ at $M_{\rm gr}=1.4~M_{\odot}$ is also noted.}
	\end{table}
	
	We found that in the case of the maximum gravitational mass, thermal effects are negligible. In particular, while the introduction of temperature ($T=1~{\rm MeV}$) leads to a lower maximum gravitational mass than the cold neutron star, the increase of temperature leads to an increasing behavior of the maximum gravitational mass. The above results confirm similar studies concerning thermal effects on the maximum neutron star mass (Nicotra et al.~\citeyear{AA.451.1.2010}; Burgio et al.~\citeyear{Burgio2007}; Burgio \& Schulze~\citeyear{AA.518.A17.2010}; Lu et al.~\citeyear{PhysRevC.100.054335}; Figura et al.~\citeyear{PhysRevD.102.043006}). However, thermal effects appear to be more important for the radius of neutron stars. For a neutron star with $M_{\rm gr}=1.4~M_{\odot}$ the radius can reach values even twice the radius of the cold one. It is worth noting that after $T=20~{\rm MeV}$, there are no configurations for a neutron star with $M_{\rm gr}=1.4~M_{\odot}$. Moreover, the maximum baryon mass decreases with increasing temperature up to $T=40~{\rm MeV}$, while for higher temperatures, a relatively low increase is observed. We concluded that hot neutron stars can exist with maximum baryon masses at lower values compared to the cold ones. These bulk properties are summarized in Table~\ref{tab:nonrotating_data}. It has to be noted here that in the case of a very hot neutron star ($T=60~{\rm MeV}$), the central density is $\sim 24\%$ lower compared to the cold case. The reason is that while the gravitational masses are comparable, the corresponding radius at $T=60~{\rm MeV}$ is $\sim 35\%$ higher than the cold case. In particular, at higher temperatures, thermal pressure, which added to the baryonic one, becomes appreciable and pushes the neutron star matter against gravity. In this case, while gravitational mass is almost unaffected, as it is mainly determined by the high-density behavior of the EOS, the radius of the star, which is determined by the low- and intermediate-density domain of the EOS, increases appreciably. As a result, the central baryon density of a hot neutron star decreases compared to the cold one.
	
	\begin{table}
		\footnotesize
		\caption{Summary of Nonrotating Isentropic Neutron Star Bulk Properties}
		\begin{tabular*}{\columnwidth}{@{\extracolsep{\fill} }  l  c  c  c  c  c  r  c }
			\toprule
			\toprule
			$Y_{l}$ & $S$ & $M_{b}^{\rm max}$ & $M_{gr}^{\rm max}$ & $R_{\rm max}$ & $n_{b}^{c}$ & $T_{c}~\text{}~\text{ }$ & $R_{1.4}$ \\
			 & $(k_{B})$ & $(M_{\odot})$ & $(M_{\odot})$ & (km) & ${\rm (fm^{-3})}$ & (MeV) & (km) \\
			\midrule
			\decimals
			{} & 1 & 2.612 & 2.196 & 10.678 & 1.049 & 31.5 & 12.384 \\
			{0.2} & 2 & 2.589 & 2.213 & 11.335 & 0.946 & 66.2 & 13.744 \\
			{} & 3 & 2.530 & 2.251 & 12.188 & 0.913 & 129.7 & 17.749 \\
			\hline
			{} & 1 & 2.515 & 2.149 & 10.678 & 1.075 & 29.6 & 12.920 \\
			{0.3} & 2 & 2.485 & 2.161 & 11.103 & 1.037 & 63.5 & 14.303 \\
			{} & 3 & 2.440 & 2.192 & 12.141 & 0.941 & 108.3 & 18.305 \\
			\hline
			{} & 1 & 2.430 & 2.110 & 10.712 & 1.110 & 28.5 & 13.679 \\
			{0.4} & 2 & 2.398 & 2.120 & 11.154 & 1.071 & 59.9 & 15.316 \\
			{} & 3 & 2.354 & 2.147 & 12.208 & 0.972 & 97.8 & 19.922  \\
			\bottomrule
		\end{tabular*}
		\label{tab:nonrotating_data_isentropic}
		\tablecomments{\footnotesize Reported are the lepton fraction $Y_{l}$, entropy per baryon $S$, baryon mass $M_{b}^{\rm max}$, gravitational mass $M_{\rm gr}^{\rm max}$, equatorial radius $R_{\rm max}$, central baryon density $n_{b}^{c}$, and central temperature $T_{c}$. The above properties correspond to the maximum gravitational mass configuration. The equatorial radius $R_{1.4}$ at $M_{\rm gr}=1.4~M_{\odot}$ is also noted.}
	\end{table}

	By considering an isentropic EOS, the mentioned quantities alter in correspondence to lepton fraction and entropy per baryon. In particular, we compare EOSs with constant lepton fractions. The increase of the entropy per baryon in neutron stars leads to lower baryon masses, as well as lower central baryon densities. In contrast to these quantities, the maximum gravitational mass, the corresponding equatorial radius, and the central temperature are increasing as the entropy per baryon increases. As the center of the star becomes hotter with increasing entropy per baryon, the baryon mass that it can withstand is lower. Last but not least, for neutron stars with $M_{\rm gr}=1.4~M_{\odot}$, the radius is increasing, where for $S=3$, it can be 61$\%$ greater than the $R_{1.4}$ of the cold configuration. These bulk properties are summarized in Table~\ref{tab:nonrotating_data_isentropic}.
	
	Finally, it is worth pointing out that the maximum gravitational/baryon mass as a function of temperature presents a strong dependence on the nuclear EOS (Sumiyoshi et al.~\citeyear{refId0}; Kaplan et al.~\citeyear{Kaplan_2014}; da Silva Schneider et al.~\citeyear{Schneider_2020}; Raduta et al.~\citeyear{10.1093/mnras/staa2491}).
	
	\subsection{Thermal Effects on Rotating Neutron Stars} \label{sec:thermal effects rotating}
	
	\begin{table*}
		\footnotesize
		\caption{Summary of Uniformly Rotating Isothermal Neutron Star Bulk Properties at the Mass-shedding Limit}
		\begin{tabular*}{\textwidth}{@{\extracolsep{\fill} }  l  c  c  c  c  c  c  c  c  c }
			\toprule
			\toprule
			$T$ & $M_{b}^{\rm max}$ & $M_{gr}^{\rm max}$ & $R_{\rm max}$ & $n_{b}^{c}$ & $R_{1.4}$ & $f_{\rm max}$ & $\mathcal{K}_{\rm max}$ & $I_{\rm max}$ & $(T/W)_{\rm max}$ \\
			(MeV) & $(M_{\odot})$ & $(M_{\odot})$ & (km) & ${\rm (fm^{-3})}$ & (km) & (Hz) &  & $(10^{38}~{\rm kg~m^{2}})$ & $(10^{-1})$ \\
			\midrule
			\decimals
			0 & 3.085 & 2.623 & 14.292 & 0.927 & 17.413 & 1689 & 0.692 & 3.949 & 1.299 \\
			1 & 2.983 & 2.549 & 13.299 & 0.935 & 15.504 & 1613 & 0.647 & 3.651 & 1.137 \\
			2 & 2.899 & 2.508 & 13.848 & 0.877 & 16.446 & 1525 & 0.622 & 3.597 & 1.055 \\
			5 & 2.730 & 2.419 & 13.887 & 0.879 & 18.501 & 1384 & 0.545 & 3.227 & 0.805 \\
			10 & 2.593 & 2.364 & 13.911 & 0.934 & 22.520 & 1285 & 0.478 & 2.914 & 0.612 \\
			15 & 2.514 & 2.338 & 14.340 & 0.934 & 26.895 & 1205 & 0.442 & 2.813 & 0.521 \\
			20 & 2.485 & 2.348 & 17.011 & 0.911 & 37.543 & 1226 & 0.456 & 2.875 & 0.549 \\
			30 & 2.427 & 2.336 & 18.079 & 0.883 & - & 1123 & 0.422 & 2.873 & 0.470 \\
			40 & 2.427 & 2.347 & 18.425 & 0.873 & - & 1090 & 0.416 & 2.949 & 0.460 \\
			50 & 2.439 & 2.365 & 18.920 & 0.848 & - & 1053 & 0.414 & 3.084 & 0.460 \\
			60 & 2.496 & 2.403 & 19.793 & 0.762 & - & 1003 & 0.433 & 3.490 & 0.531 \\
			\bottomrule
		\end{tabular*}
		\label{tab:maximally_rotating_data}
		\tablecomments{\footnotesize Reported are the temperature $T$, baryon mass $M_{b}^{\rm max}$, gravitational mass $M_{\rm gr}^{\rm max}$, equatorial radius $R_{\rm max}$, central baryon density $n_{b}^{c}$, frequency $f_{\rm max}$, Kerr parameter $\mathcal{K}_{\rm max}$, moment of inertia $I_{\rm max}$, and ratio of rotational kinetic to gravitational binding energy $(T/W)_{\rm max}$. The above properties correspond to the maximum gravitational mass configuration. The equatorial radius $R_{1.4}$ at $M_{\rm gr}=1.4~M_{\odot}$ is also noted.}
	\end{table*}

	\begin{figure}
		\includegraphics[width=\columnwidth]{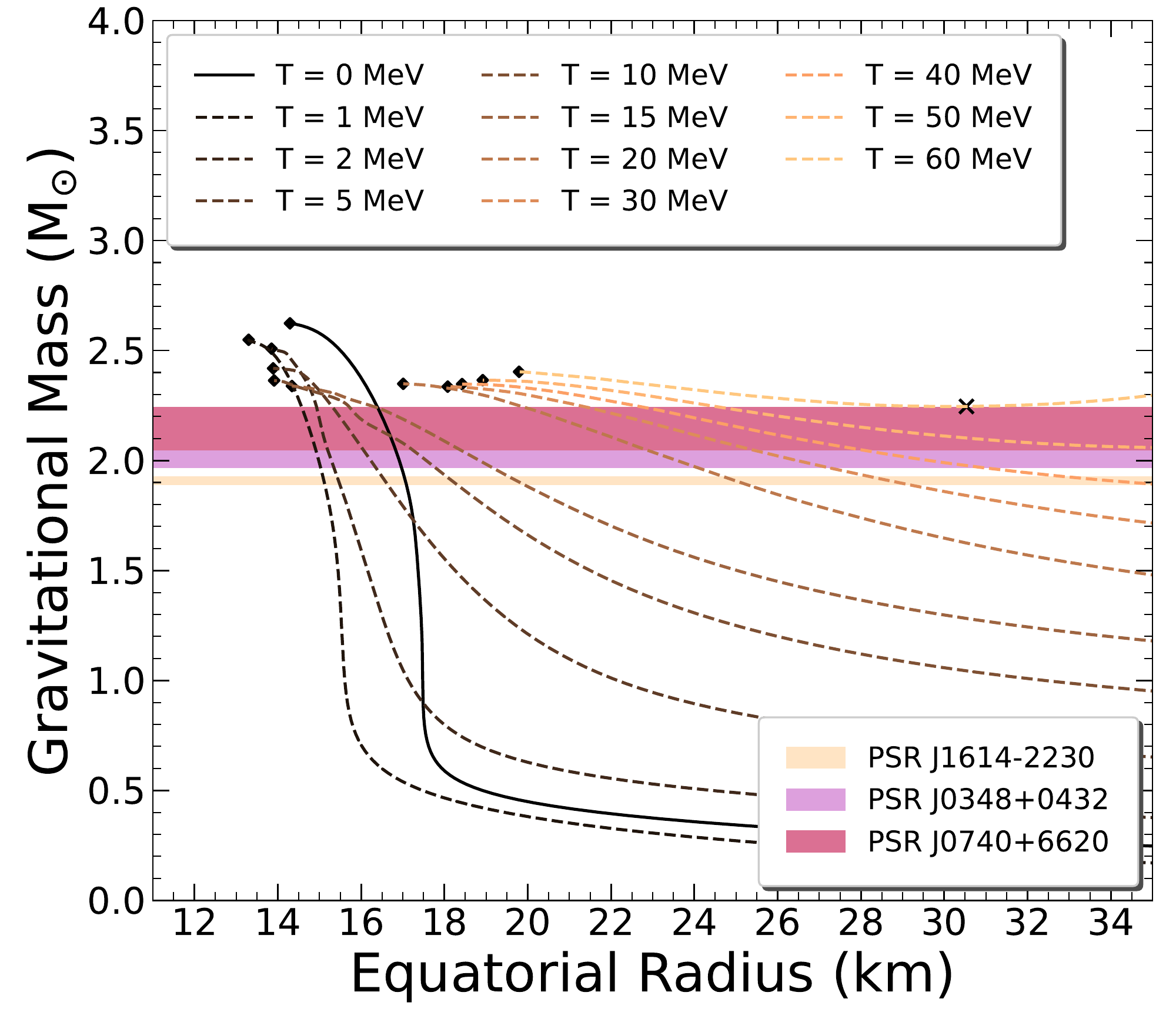}
		\caption{Gravitational mass as a function of equatorial radius for temperatures in the range $[0,60]~{\rm MeV}$ at the rotating configuration with Kepler frequency. The cold configuration is presented by the solid line, while the hot configurations are presented by the dashed ones. The shaded regions from bottom to top represent the PSR J1614-2230 (Arzoumanian et al.~\citeyear{Arzoumanian-2018}), PSR J0348+0432 (Antoniadis et al.~\citeyear{Antoniadis-2013}), and PSR J0740+6620 (Cromartie et al.~\citeyear{Cromartie-2019}) pulsar observations with possible maximum neutron star mass. Black diamonds correspond to the maximum mass configuration in each case, while the black cross corresponds to the minimum mass configuration (the remaining minimum masses are positioned at higher values of equatorial radius).}
		\label{fig:mass radius kepler}
	\end{figure}

	In Figure~\ref{fig:mass radius kepler}, we display the gravitational mass as a function of the corresponding equatorial radius at the mass-shedding limit for temperatures in the range $[0,60]~{\rm MeV}$. In general, as the temperature increases, the bulk properties of neutron stars at the maximum mass configuration\footnote{At the mass-shedding limit, we consider that maximum mass corresponds to Kepler frequency (Friedman \& Stergioulas~\citeyear{friedman_stergioulas_2013}).} are affected. In particular, the dependence of the baryon mass on the temperature exhibits similar behavior with the nonrotating case. However, the gravitational mass is decreasing with increasing temperature up to $T=30~{\rm MeV}$, while for higher values of temperature, an inverse behavior is observed. Similar to the nonrotating case, while the introduction of temperature $(T=1~{\rm MeV})$ leads to a lower value of the corresponding equatorial radius than the cold neutron star, the equatorial radius follows an increasing path with the temperature, where for neutron stars with $M_{\rm gr}=1.4~M_{\odot}$, it can reach several times the radius of the cold one with a dramatic increase. These results play a significant role in the time evolution of hot and rapidly rotating neutron stars. The temperature dependence of the maximum gravitational mass and corresponding equatorial radius is well reflected on the corresponding temperature dependence of the rest of the neutron star properties, including the central baryon density, Kepler frequency, Kerr parameter, moment of inertia, and ratio of rotational kinetic to gravitational binding energy, as displayed in Table~\ref{tab:maximally_rotating_data}.
	
	In the case of isentropic EOSs, the increase of entropy per baryon affects the neutron star bulk properties in the maximum mass configuration at the mass-shedding limit. To be more specific, considering a constant lepton fraction, the bulk properties under consideration are decreasing as the entropy per baryon; consequently, the temperature at the center of the star increases. Exceptionally, the equatorial radius follows the opposite direction, as it is increasing with the entropy per baryon. This effect is more pronounced at $M_{\rm gr} = 1.4~M_{\odot}$, where the radius can rise up to 49$\%$ of the cold star. These bulk properties are summarized in Table~\ref{tab:maximally_rotating_data_isentropic}.
	
	\begin{table*}
		\footnotesize
		\caption{Summary of Uniformly Rotating Isentropic Neutron Star Bulk Properties at the Mass-shedding Limit}
		\begin{tabular*}{\textwidth}{@{\extracolsep{\fill} }  l  c  c  c  c  c  r  c  c  c  c  c }
			\toprule
			\toprule
			$Y_{l}$ & $S$ & $M_{b}^{\rm max}$ & $M_{gr}^{\rm max}$ & $R_{\rm max}$ & $n_{b}^{c}$ & $T_{c}~\text{}~\text{ }$ & $R_{1.4}$ & $f_{\rm max}$ & $\mathcal{K}_{\rm max}$ & $I_{\rm max}$ & $(T/W)_{\rm max}$ \\
			 & $(k_{B})$ & $(M_{\odot})$ & $(M_{\odot})$ & (km) & ${\rm (fm^{-3})}$ & (MeV) & (km) & (Hz) &  & $(10^{38}~{\rm kg~m^{2}})$ & $(10^{-1})$ \\
			\midrule
			\decimals
			{} & 1 & 3.050 & 2.599 & 14.028 & 0.958 & 29.6 & 17.498 & 1715 & 0.684 & 3.775 & 1.269 \\
			{0.2} & 2 & 2.954 & 2.560 & 14.621 & 0.926 & 65.2 & 19.415 & 1594 & 0.641 & 3.694 & 1.107 \\
			{} & 3 & 2.808 & 2.517 & 15.970 & 0.879 & 125.3 & 25.912 & 1391 & 0.568 & 3.621 & 0.857 \\
			\hline
			{} & 1 & 2.817 & 2.431 & 12.780 & 0.979 & 27.9 & 15.770 & 1575 & 0.601 & 3.158 & 0.976 \\
			{0.3} & 2 & 2.743 & 2.407 & 13.314 & 0.940 & 59.5 &  17.590 & 1458 & 0.565 & 3.143 & 0.861 \\
			{} & 3 & 2.633 & 2.380 & 14.116 & 0.908 & 105.5 & 22.861 & 1261 & 0.493 & 3.104 & 0.651 \\
			\hline
			{} & 1 & 2.733 & 2.398 & 13.906 & 1.052 & 27.5 & 19.454 & 1661 & 0.613 & 2.972 & 1.005 \\
			{0.4} & 2 & 2.659 & 2.371 & 14.702 & 1.002 & 57.5 & 21.975 & 1530 & 0.576 & 2.965 & 0.885 \\
			{} & 3 & 2.519 & 2.309 & 14.185 & 0.943 & 95.6 & 25.020 & 1225 & 0.466 & 2.841 & 0.580 \\
			\bottomrule
		\end{tabular*}
		\label{tab:maximally_rotating_data_isentropic}
		\tablecomments{\footnotesize Reported are the lepton fraction $Y_{l}$, entropy per baryon $S$, baryon mass $M_{b}^{\rm max}$, gravitational mass $M_{\rm gr}^{\rm max}$, equatorial radius $R_{\rm max}$, central baryon density $n_{b}^{c}$, central temperature $T_{c}$, frequency $f_{\rm max}$, Kerr parameter $\mathcal{K}_{\rm max}$, moment of inertia $I_{\rm max}$, and ratio of rotational kinetic to gravitational binding energy $(T/W)_{\rm max}$. The above properties correspond to the maximum gravitational mass configuration. The equatorial radius $R_{1.4}$ at $M_{\rm gr}=1.4~M_{\odot}$ is also noted.}
	\end{table*}
	
	\begin{table*}
		\footnotesize
		\caption{Minimum Mass of Isothermal and Isentropic Neutron Stars}
		\begin{tabular*}{\textwidth}{@{\extracolsep{\fill} }  l  c  c  c  c  c  c  c  c  c  c  c }
			\toprule
			\toprule
			$M_{\rm gr}^{\rm min}$ & $T=0$ & $T=1$ & $T=2$ & $T=5$ & $T=10$ & $T=15$ & $T=20$ & $T=30$ & $T=40$ & $T=50$ & $T=60$ \\
			$(M_{\odot})$ & $(\rm MeV)$ & $(\rm MeV)$ & $(\rm MeV)$ & $(\rm MeV)$ & $(\rm MeV)$ & $(\rm MeV)$ & $(\rm MeV)$ & $(\rm MeV)$ & $(\rm MeV)$ & $(\rm MeV)$ & $(\rm MeV)$ \\
			\midrule
			\decimals
			N.R. & 0.080 & 0.098 & 0.172 & 0.353 & 0.652 & 0.893 & 1.116 & 1.471 & 1.720 & 1.888 & 2.003 \\
			M.R. & 0.081 & 0.098 & 0.172 & 0.356 & 0.671 & 0.915 & 1.142 & 1.545 & 1.848 & 2.053 & 2.242 \\
			\midrule
			\midrule
			{} & \multicolumn{3}{c}{$Y_{l}=0.2$} & \multicolumn{3}{c}{$Y_{l}=0.3$} & \multicolumn{3}{c}{$Y_{l}=0.4$} \\
			& $S=1$ & $S=2$ & $S=3$ & $S=1$ & $S=2$ & $S=3$ & $S=1$ & $S=2$ & $S=3$ \\
			\midrule
			& 0.172 & 0.270 & 0.569 & 0.285 & 0.410 & 0.670 & 0.426 & 0.559 & 0.821
\\
			& 0.186 & 0.318 & 0.590 & 0.320 & 0.417 & 0.683 & 0.432 & 0.569 & 0.823 \\
			\bottomrule
		\end{tabular*}
		\label{tab:minimum_mass}
		\tablecomments{\footnotesize Reported are the temperature $T$, lepton fraction $Y_{l}$, entropy per baryon $S$ in units of $k_{B}$, and gravitational mass $M_{\rm gr}^{\rm min}$. The abbreviation ``N.R." correspronds to the nonrotating configuration, while the ``M.R." corresponds to the maximally rotating one.}
	\end{table*}
	
	\subsection{Minimum Mass of Neutron Stars}
	Apart from the maximum neutron star mass, the minimum one is also of great interest in astrophysics. For reasons of completeness, we study the thermal and rotation effects on the minimum mass of neutron stars. The existence of a minimum neutron star configuration is a universal feature, independent of the details of the EOS (Colpi et al.~\citeyear{Colpi_1989}), for example, the concept of the minimum mass involved in the case of a neutron star in a close binary system with a more compact partner (neutron star or black hole; Suwa et al.~\citeyear{10.1093/mnras/sty2460}). During evolution, the lower-mass neutron star transfers mass to the more massive object, a process that ultimately leads to approaching its minimum value. Finally, crossing this value, the neutron star reaches a nonequilibrium configuration (Haensel et al.~\citeyear{refId1}). In particular, it is pointed out by several authors that a neutron star will undergo an explosion if its mass drops to the minimum possible equilibrium value~ (Blinnikov et al.~\citeyear{Blinnikov_1984}; Colpi et al.~\citeyear{Colpi_1989}, ~\citeyear{Colpi_1991}; Sumiyoshi et al.~\citeyear{Sumiyoshi_1998}). In most studies, the minimum mass is studied in the framework of cold catalyzed nuclear matter. In the present work, we extend previous studies in order to include rotation and thermal effects, which are related to a more realistic process. The results are displayed in Table~\ref{tab:minimum_mass}.
		
	In the case of isothermal configurations, the increase in temperature leads to a significant increase of the minimum mass, especially for high values of temperature. On the other hand, the rotation effect is important only for high-temperature configurations. The latter occurs because of the low values of Kepler frequency at low temperatures. In this case, the difference in minimum mass between the nonrotating and maximally rotating configurations is almost imperceptible.
	
	Similarly, in adiabatic cases, higher values of entropy per baryon lead to higher values of minimum mass (for a constant lepton fraction). Moreover, for constant entropy per baryon, neutron stars that are rich in leptons exhibit higher values of minimum mass. However, the most distinctive feature in isentropic configurations is the negligible effect of the rotation on the minimum mass, in most of the cases. The explanation of this behavior is similar to that of isothermal cases, that is, the low corresponding values of Kepler frequency.
	
	\subsection{Sequences of Constant Baryon Mass and the Threshold Mass of Cold, Catalyzed Neutron Stars}
	
	\begin{figure}
		\includegraphics[width=\columnwidth]{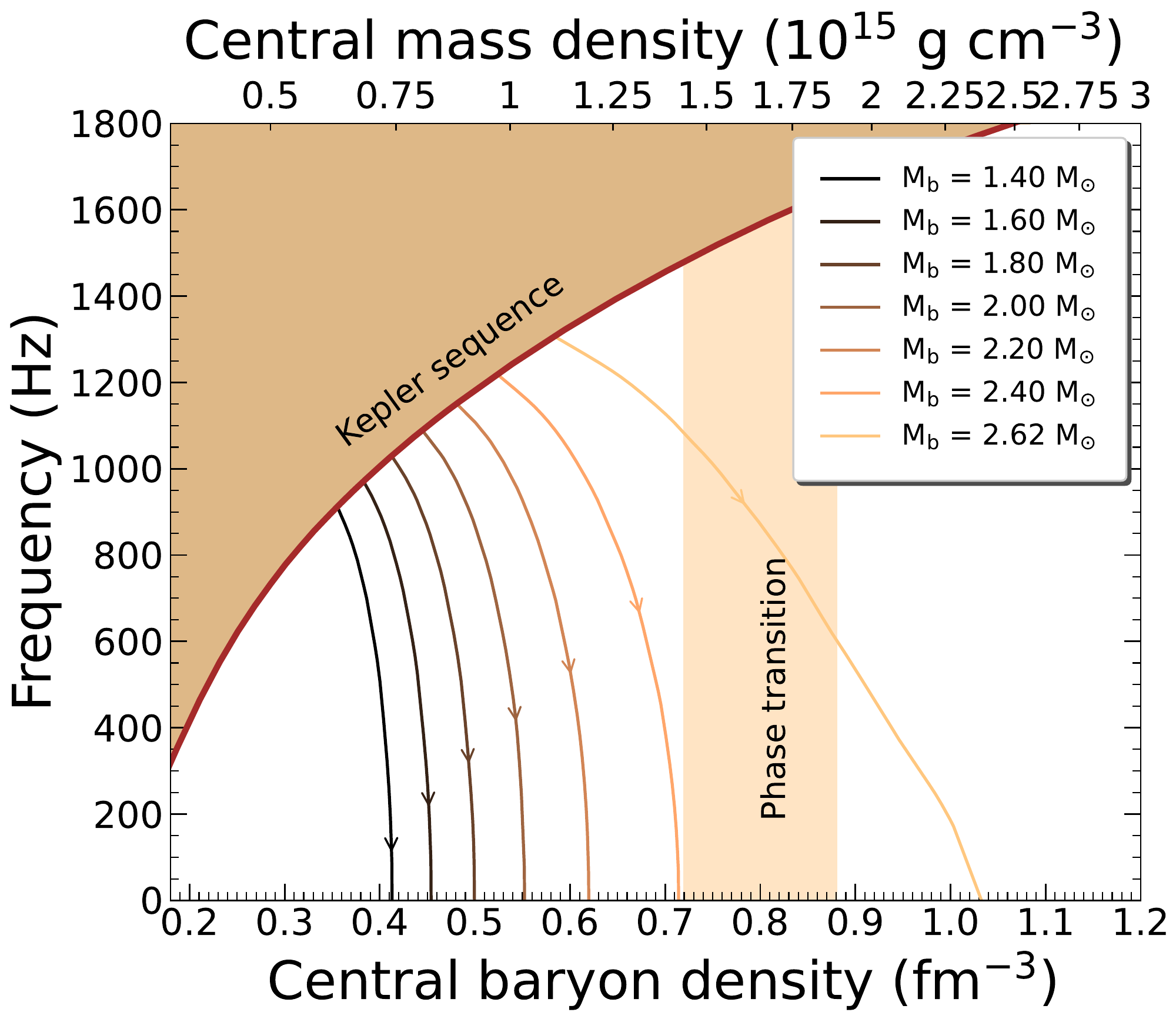}
		\caption{Frequency as a function of central stellar baryon density for constant baryon mass sequences. The shaded region represents the forbidden region for the star, where the boundary solid line marks the Kepler frequency. Arrows are shown to guide the evolution of the star. A region where a possible phase transition may occur is also noted (Baym et al.~\citeyear{Baym_2018}). In addition, the central mass density is presented in the top axis corresponding to central baryon density.}
		\label{fig:frequency vs density cold}
	\end{figure}
	
	In Figure~\ref{fig:frequency vs density cold}, we display sequences of constant baryon mass up to the one that corresponds to the maximum gravitational mass configuration in the case of cold catalyzed matter. From these sequences, it is clear that, differently from the gravitational mass where changes are negligible, as the frequency decreases, starting from the Kepler frequency, the star gets considerably more dense. The effect reaches its peak for baryon masses close to the one that corresponds to the maximum gravitational mass configuration ($M_{b} = 2.62\ M_{\odot}$), and it will be reflected in the particle composition and thermal properties. In addition, we have indicated the region where a possible phase transition may occur ($0.72~{\rm fm^{-3}} \leq n_{\rm tr} \leq 0.88~{\rm fm^{-3}}$; Baym et al.~\citeyear{Baym_2018}). The results of Baym et al.~\citeyearpar{Baym_2018} are only indicative and simply provide a possible region of transition density (from baryonic to quark matter). Other similar studies predict similar or different corresponding regions. Obviously, more robust (theoretical and experimental) constraints concerning the phase transition are needed.
	
	In this case, Figure~\ref{fig:frequency vs density cold} may help to indicate the expected region of the central densities (for a constant baryon mass) where a possible phase transition may take place during the evolution of a neutron star. In particular, this study may be useful for the evolution of pulsars and the appearance of the back-bending process (Glendenning~\citeyear{Glendenning-2000}).
	
	Finally, for a given cold, catalyzed EOS, one can define the threshold binary mass that distinguishes the prompt ($M_{\rm st}^{\rm max} > M_{\rm thres}$) from the delayed ($M_{\rm st}^{\rm max} < M_{\rm thres}$) collapse. A relation that describes the threshold mass as a function of the compactness was found recently in K{\"o}ppel et al.~\citeyearpar{K_ppel_2019} and is given by
	\begin{equation}
		M_{\rm thres} = M_{\rm st}^{\rm max} \left(3.06-\frac{1.01}{1-1.34\beta_{\rm max}}\right),
		\label{eq:threshold mass}
	\end{equation}
	where $\beta$ is the compactness parameter of the star, defined as
	\begin{equation}
		\beta = \frac{G}{c^{2}}\frac{M}{R},
		\label{eq:compactness}
	\end{equation}

	\begin{figure*}
		\includegraphics[width=\textwidth]{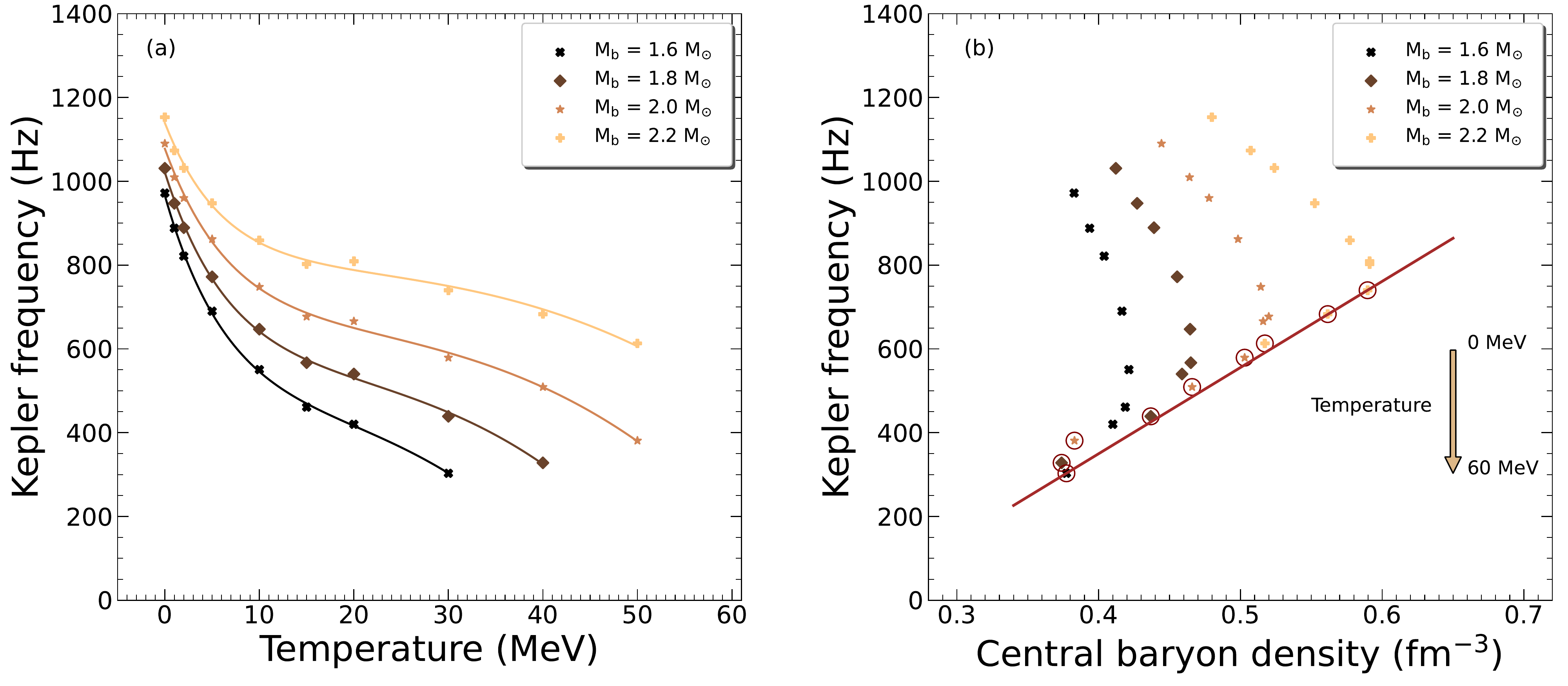}
		\caption{Kepler frequency as a function of (a) temperature and (b) central baryon density for constant baryon mass sequences. (a) Solid lines correspond to fits originated from Equation~\eqref{eq_s9_6:fit_f_t_lt}. (b) The solid line corresponds to Equation~\eqref{eq_s9_6:fit_f_nb_trend} and open circles mark the high-temperature region ($T\geq 30$ MeV).}
		\label{fig:frequency vs temperature and density}
	\end{figure*}
	
	\noindent and $\beta_{\rm max}$ corresponds to the maximum mass configuration. In our case, employing the values of $M_{\rm st}^{\rm max}$ and $\beta_{\rm max}$, we found that $M_{\rm thres} = 2.994~M_{\odot}$. Although the remnant is expected to rotate differentially and not uniformly, we present the threshold mass in this study in order to show that uniform rotation cannot reach the values of gravitational and baryon mass, as Table~\ref{tab:maximally_rotating_data} indicates. The implementation of differential rotation will be the subject of a forthcoming paper.
	
	\subsection{Sequences of Constant Baryon Mass on Rotating Neutron Stars at Finite Temperature}

	The sequences of constant baryon mass are a very useful way to study thermal effects on the evolution and instability conditions of hot neutron stars. However, as we have constructed isothermal EOSs, we studied the same baryon mass configuration in the temperature range $[0,60]~{\rm MeV}$ and eventually constructed a sequence related to the cooling of a neutron star. In particular, the quantities under consideration were the Kepler frequency, the central baryon density, and the temperature of each EOS.
	
	Figure~\ref{fig:frequency vs temperature and density}(a) displays the Kepler frequency as a function of temperature for four baryon masses. As the temperature increases, the Kepler frequency presents a reverse behavior. More specifically, while until $T=15~{\rm MeV}$, the reduction of the Kepler frequency is rather abrupt, for higher temperatures, a smoother one is observed. The dependence of the Kepler frequency on the temperature is described by the formula
	\begin{equation}
		f(T) = a_{0} + a_{1} T^{3} + a_{2} \exp[a_{3} T] \quad (\rm Hz),
		\label{eq_s9_6:fit_f_t_lt}
	\end{equation}
	where $f$ and $T$ are given in units of $\rm Hz$ and $\rm MeV$, respectively, and the coefficients $a_{i}$, with $i=0-3$, are presented in Table~\ref{tab:fit1+2}.
	
	In addition, Figure~\ref{fig:frequency vs temperature and density}(b) displays the Kepler frequency as a function of the central baryon density for four baryon masses. The central baryon density presents exceptional behavior in that as the temperature increases, the central baryon density is also increased, but for high values of temperature, it exhibits an inverse behavior. In any case, these effects are mild. However, it is worth noticing that the corresponding effect is sizable, leading to a reduction of two to three times the Kepler frequency. Furthermore, the most distinctive feature is the appearance of an almost linear relation between the Kepler frequency and the central baryon density for a constant value of temperature, especially for low ones. Moreover, and quite notably, we found that for high values of temperature $(T\geq 30~{\rm MeV})$, every sequence of constant baryon mass not only presents similar behavior but also moves along a linear relation described as
	\begin{equation}
		f(n_{b}^{c}) = -473.144 + 2057.271n_{b}^{c} \quad (\rm Hz),
		\label{eq_s9_6:fit_f_nb_trend}
	\end{equation}
	where $f$ and $n_{b}^{c}$ are given in units of Hz and $\rm fm^{-3}$, respectively. Equation~\eqref{eq_s9_6:fit_f_nb_trend} is very useful, since it directly relates the Kepler frequency with the central baryon density of a very hot neutron star, independently of the corresponding  baryon mass. In addition, this relation defines the allowed region for rotation with the Kepler frequency of a hot neutron star for a specific value of the central baryon density, and vice versa.
	
	\begin{table}
		\footnotesize
		\caption{Coefficients of Empirical Relations~\eqref{eq_s9_6:fit_f_t_lt} and~\eqref{eq_s9_6:fit_nb_t_lt} for Baryon Masses in the Range $[1.6-2.2]~M_{\odot}$}
		\begin{tabular*}{\columnwidth}{@{\extracolsep{\fill} }  l  r  r  r r }
			\toprule
			\toprule
			Coefficients & \multicolumn{4}{c}{Baryon Mass} \\
			\cline{2-5}
			 & $1.6~M_{\odot}$ & $1.8~M_{\odot}$ & $2.0~M_{\odot}$ & $2.2~M_{\odot}$\\
			 \midrule
			 $a_{0}~(\times 10^{2})$ & 4.259 & 5.284 & 6.414 & 7.863 \\
			 $a_{1}~(\times 10^{-3})$ & -4.787 & -3.202 & -2.099 & -1.443 \\
			 $a_{2}~(\times 10^{2})$ & 5.401 & 4.929 & 4.363 & 3.530 \\
			 $a_{3}~(\times 10^{-1})$ & -1.468 & -1.443 & -1.424 & -1.636 \\
			 $b_{0}~(\times 10^{-1})$ & 4.273 & 4.466 & 4.798 & 5.470 \\
			 $b_{1}~(\times 10^{-2})$ & -0.075 & 0.638 & 1.138 & 1.204 \\
			 $b_{2}~(\times 10^{-6})$ & -1.699 & -1.754 & -1.405 & -0.926 \\
			 $b_{3}~(\times 10^{-2})$ & -4.473 & -3.512 & -3.609 & -6.640 \\
			 $b_{4}~(\times 10^{-1})$ & -3.299 & -3.608 & -3.357 & -2.389 \\
			 \bottomrule
		\end{tabular*}
		\label{tab:fit1+2}	
	\end{table}

	\begin{figure}
		\includegraphics[width=\columnwidth]{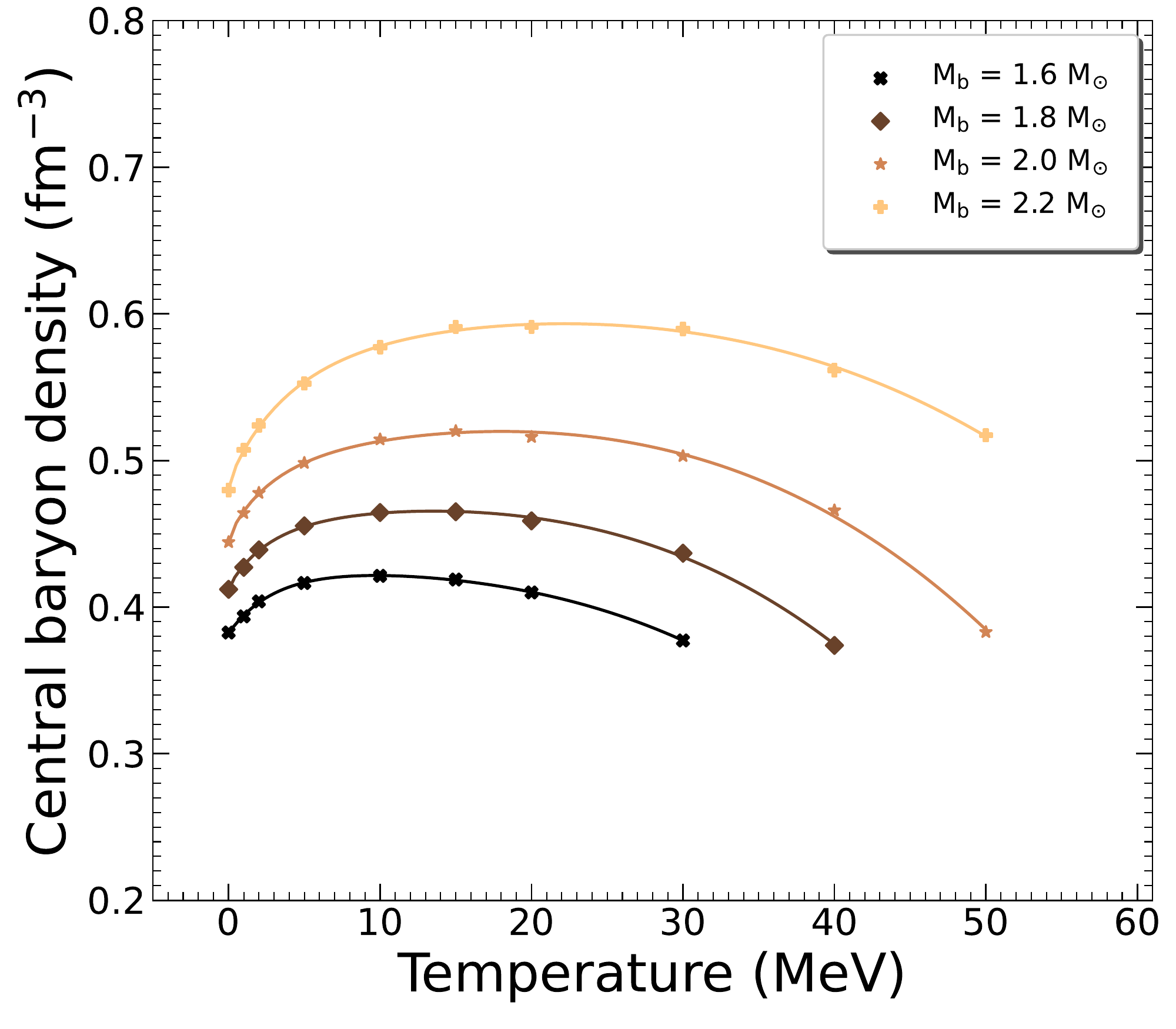}
		\caption{Central baryon density as a function of temperature for constant baryon mass sequences. Solid lines correspond to fits originated from Equation~\eqref{eq_s9_6:fit_nb_t_lt}. The configuration corresponds to the mass-shedding limit.}
		\label{fig:density vs temperature}
	\end{figure}
	
	Since it is interesting to study the dependence of the central baryon density on the temperature (for a neutron star spinning with the Kepler frequency), we provide in Figure~\ref{fig:density vs temperature} the central baryon density as a function of the temperature for four baryon masses. While for temperatures up to $T=15~{\rm MeV}$, the central baryon density is increased, for higher ones, it follows the opposite path, as its nonmonotonic behavior is presented. This behavior can be described by the formula
	\begin{equation}
		n_{b}^{c}(T) = b_{0} + b_{1} T^{1/2} + b_{2} T^{3} + b_{3} \exp[b_{4} T] \quad (\rm fm^{-3}),
		\label{eq_s9_6:fit_nb_t_lt}
	\end{equation}
	where $n_{b}^{c}$ and $T$ are given in units of $\rm fm^{-3}$ and $\rm MeV$, respectively, and the coefficients $b_{i}$, with $i=0-4$, are presented in Table~\ref{tab:fit1+2}.
	
	It has to be noted that for a given value of baryon mass, the stability range of a neutron star is defined in a specific temperature range. This is the reason why, in the corresponding figures, there are no configurations for some temperatures and baryon masses.
	
	\subsection{Moment of Inertia, Kerr Parameter, and Ratio T/W on Rotating Neutron Stars}
	
	\begin{figure}
		\includegraphics[width=\columnwidth]{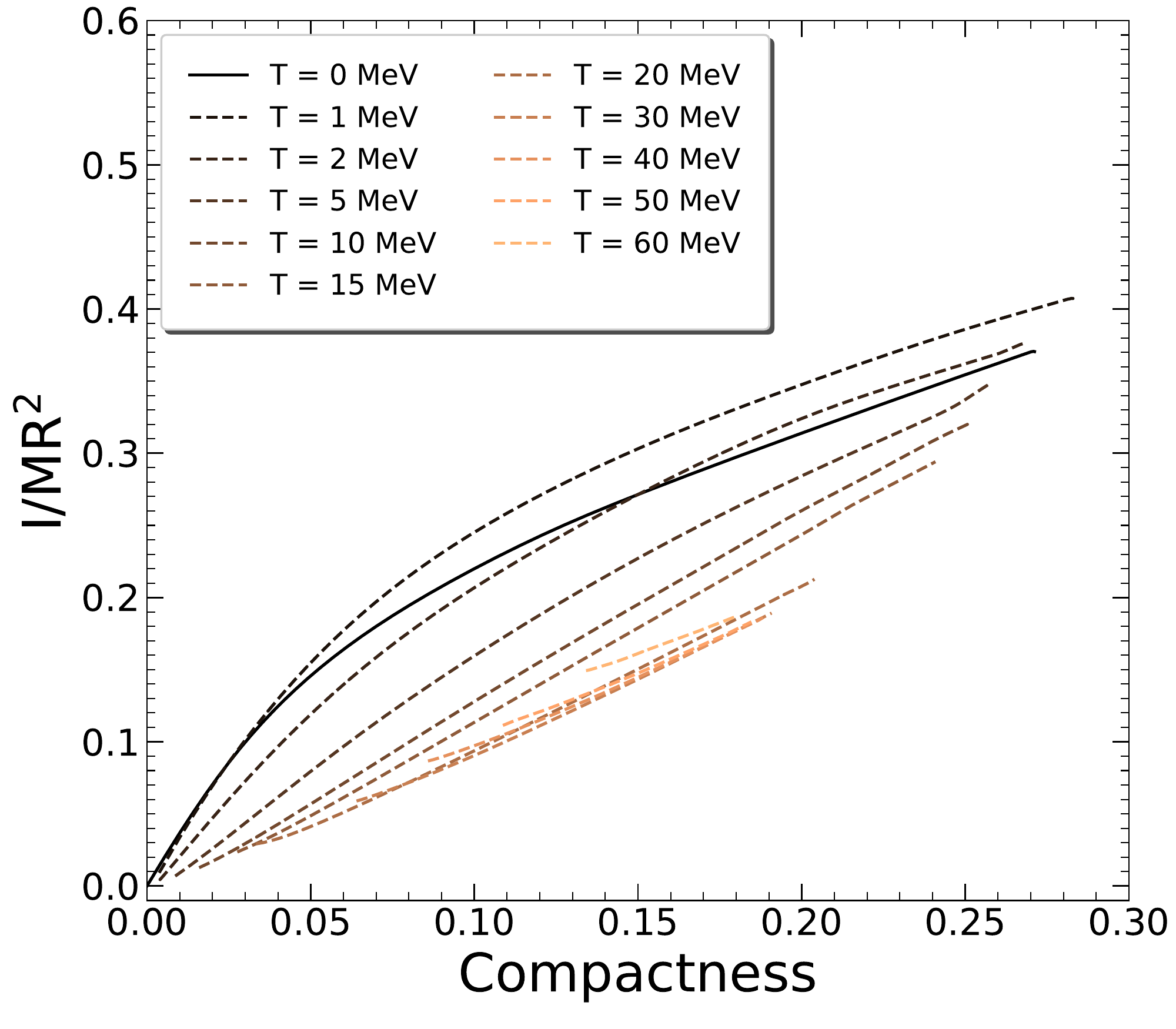}
		\caption{Dimensionless moment of inertia as a function of compactness parameter for temperatures in the range $[0,60]$ MeV. The cold configuration is presented by the solid line, while hot configurations are presented by the dashed ones. The configuration corresponds to the mass-shedding limit.}
		\label{fig:moment of inertia vs compactness}
	\end{figure}
	
	The study of rotating neutron stars offers much more information concerning the EOS compared to nonrotating ones. In the present work, we focused on studying the moment of inertia, the Kerr parameter, and the ratio of rotational kinetic to gravitational binding energy $(T/W)$ at the mass-shedding limit.

	Figure~\ref{fig:moment of inertia vs compactness} displays the dimensionless moment of inertia as a function of the compactness parameter for isothermal neutron stars. The dimensionless moment of inertia provides an important constraint for the interior structure of neutron stars. Although for low values of temperature, $T\leq 2~{\rm MeV}$, the dimensionless moment of inertia is higher than the cold neutron star, for temperatures $T>2~{\rm MeV}$, the reverse behavior is presented. This result points to the conclusion that the increase of temperature, except for some specific cases ($T < 2~{\rm MeV}$), leads to lesser compact objects than the cold neutron star.
	
	\begin{figure}
		\includegraphics[width=\columnwidth]{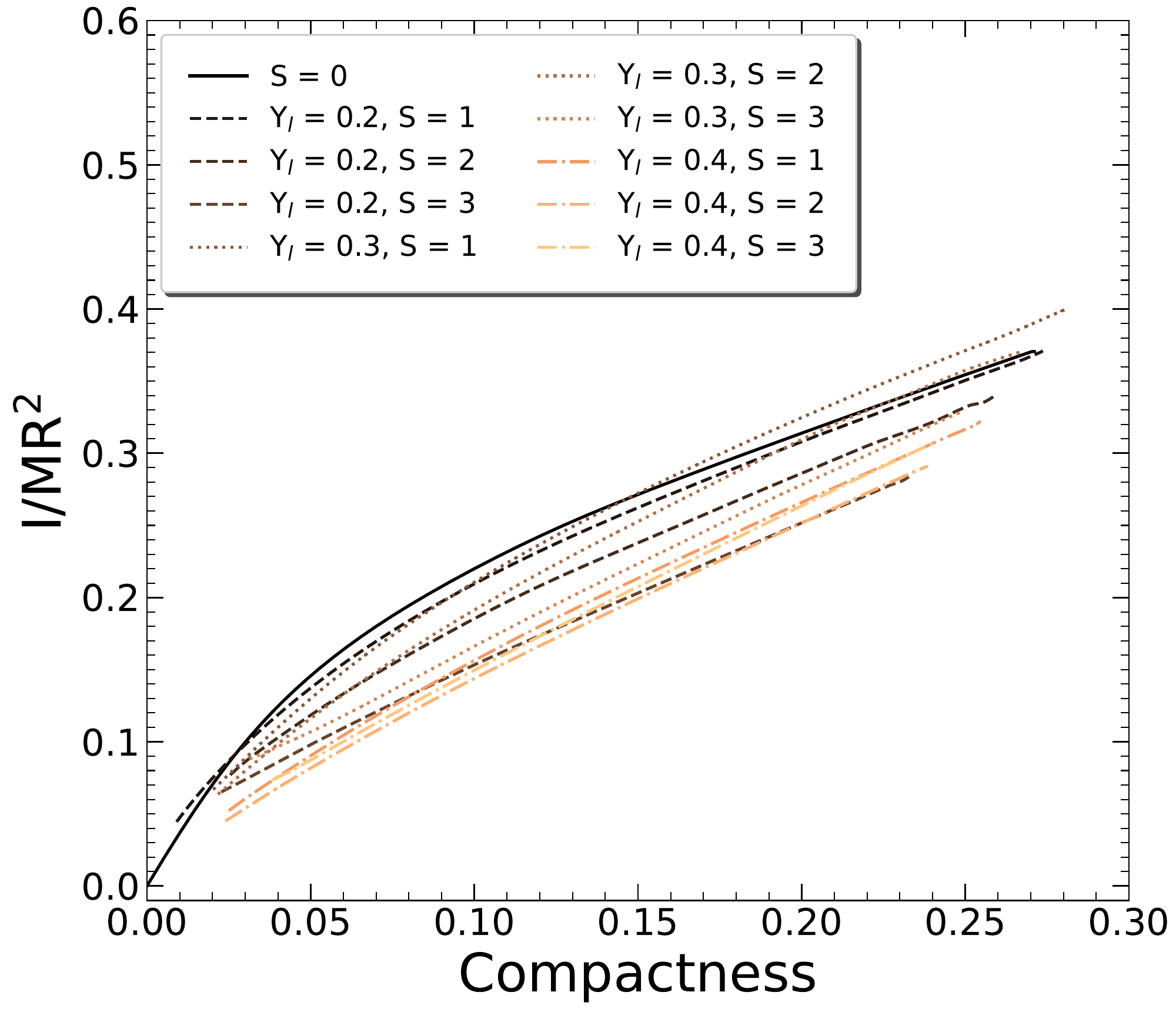}
		\caption{Dimensionless moment of inertia as a function of compactness parameter for lepton fractions and entropies per baryon in the ranges $[0.2,0.4]$ and $[1,3]~k_{B}$, respectively. The cold configuration is presented by the solid line. The configuration corresponds to the mass-shedding limit.}
		\label{fig:moment of inertia vs compactness isentropic}
	\end{figure}
	
	Figure~\ref{fig:moment of inertia vs compactness isentropic} displays the dimensionless moment of inertia as a function of the compactness parameter for isentropic neutron stars. In general, the increase of the entropy per baryon with a constant lepton fraction, leads to lesser compact objects with lower values to a dimensional moment of inertia than the cold neutron star. There are some specific cases, $Y_{l}=0.2$ and 0.3 and $S=1$, where these values exceed the limit introduced by the cold neutron star.
	
	A quantity directly related to black holes and neutron stars is the Kerr parameter (dimensionless spin parameter), which is defined as
	\begin{equation}
		\mathcal{K}\equiv \frac{c}{G}\frac{J}{M^2}=\frac{c}{G}\frac{I\Omega}{M^2}.
		\label{eq:kerr parameter}
	\end{equation}
	Its importance lies with the mass-shedding limit, where it takes the maximum allowed value. As shown in Koliogiannis \& Moustakidis~\citeyearpar{PhysRevC.101.015805}, this limit represents an indicator of the final fate of the collapse of a rotating compact star. In fact, it was found in a recent work (Koliogiannis \& Moustakidis~\citeyear{PhysRevC.101.015805}) that the Kepler angular velocity for a cold neutron star is given by an almost EOS-independent formula,
	\begin{equation}
		\Omega_{\rm k}=2\pi {\cal C}_{\rm rot}\left(\frac{M_{\rm max}^{\rm rot}}{M_{\odot}}  \right)^{1/2}\left(\frac{10 \rm km}{R_{\rm max}^{\rm rot}}  \right)^{3/2},
		\label{eq:omega kepler}
	\end{equation}
	while the moment of inertia corresponding to the Kepler frequency is given by (see also Shao et al.~\citeyear{PhysRevD.101.063029})
	\begin{equation}
		\frac{I_{\rm k}}{M_{\rm max}^{\rm rot}(R_{\rm max}^{\rm rot})^2}\simeq 1.379 \beta_{\rm max},
		\label{eq:moi kerr}
	\end{equation}
	where
	\begin{equation}
		\beta_{\rm max}=\frac{G}{c^2}\frac{M_{\rm max}^{\rm rot}}{R_{\rm \max}^{\rm rot}}.
	\end{equation}

	\begin{figure*}
		\includegraphics[width=\textwidth]{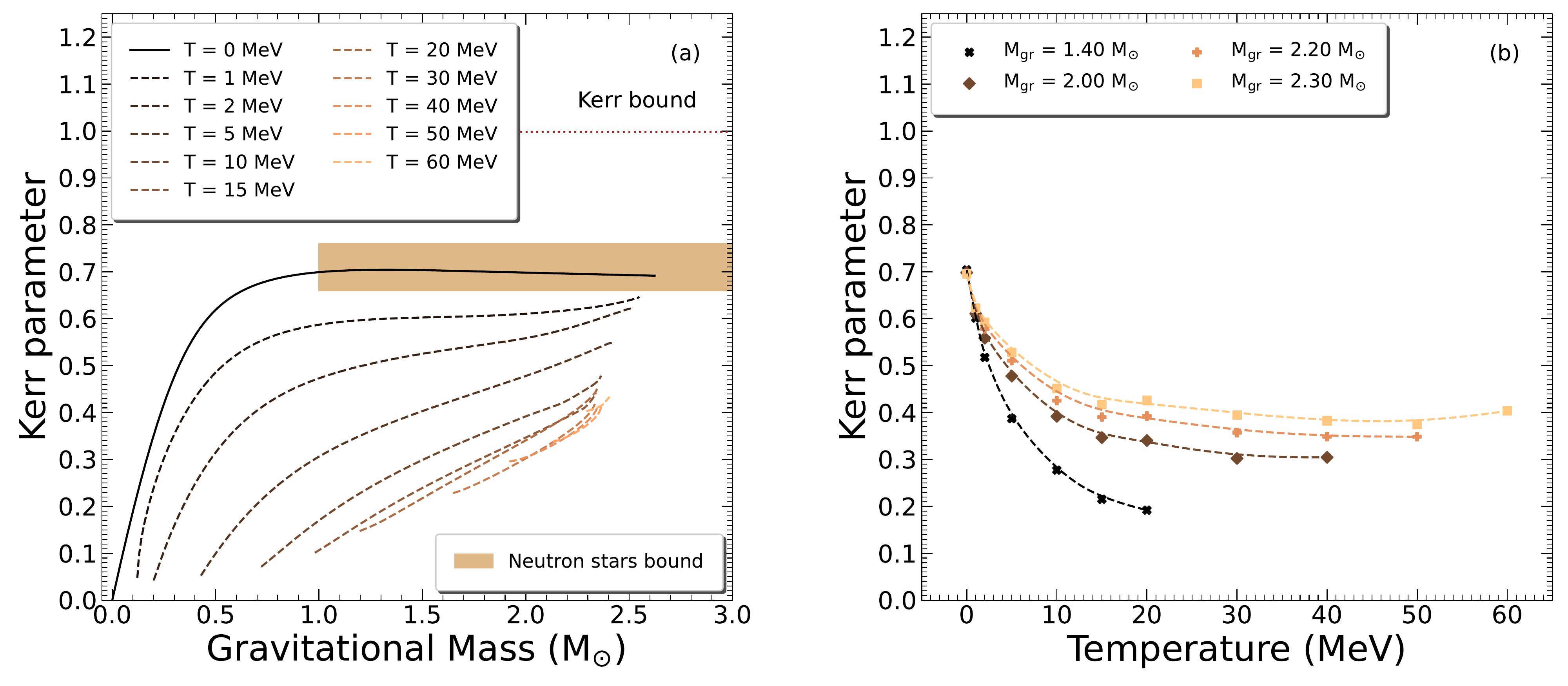}
		\caption{(a) Kerr parameter as a function of gravitational mass for temperatures in the range $[0,60]$ MeV. The horizontal dotted line marks the Kerr bound for astrophysical Kerr black holes, $\mathcal{K_{\rm B.H.}}=0.998$ (Thorne~\citeyear{Thorne_1974}). The shaded region represents the limits for neutron stars from Equation~\eqref{eq:kerr parameter 2}. The cold configuration is presented by the solid line, while the hot configurations are presented by the dashed ones. (b) Kerr parameter as a function of temperature for constant gravitational mass. The crosses represent $M_{\rm gr} = 1.4~M_{\odot}$, diamonds $M_{\rm gr} = 2~M_{\odot}$, plus signs $M_{\rm gr} = 2.2~M_{\odot}$, and squares $M_{\rm gr} = 2.3~M_{\odot}$. The configuration corresponds to the mass-shedding limit.}
		\label{fig:kerr vs mass}
	\end{figure*}

	From Equations~\eqref{eq:kerr parameter} -~\eqref{eq:moi kerr}, we found that, in a very good approximation, the Kerr parameter, at the Kepler frequency (mass-shedding limit) for a cold, catalyzed neutron star is given by the simple universal expression
	\begin{equation}
		\mathcal{K}_{\rm k}\simeq 1.34 \sqrt{\beta_{\rm max}}.
		\label{eq:kerr parameter 2}
	\end{equation}
	Considering that, for the majority of realistic cold EOSs, the relation $0.24 \leq \beta_{\rm max}  \leq 0.32$ holds, we concluded that $0.66 \leq {\cal K}_{\rm k}  \leq 0.76 $.
	
	Figure~\ref{fig:kerr vs mass}(a) displays the Kerr parameter as a function of the gravitational mass for isothermal neutron stars. The effect of the temperature has a dramatic impact on the Kerr parameter. As the temperature increases, the Kerr parameter follows a slightly decreasing trajectory, except for $T=60~{\rm MeV}$, a behavior that is also shown in Figure~\ref{fig:kerr vs mass}(b), where the Kerr parameter is plotted as a function of temperature for constant gravitational masses. It has to be noted that after $T=30~{\rm MeV}$, the Kerr parameter creates a plate for each gravitational mass configuration.
	
	\begin{figure}
		\includegraphics[width=\columnwidth]{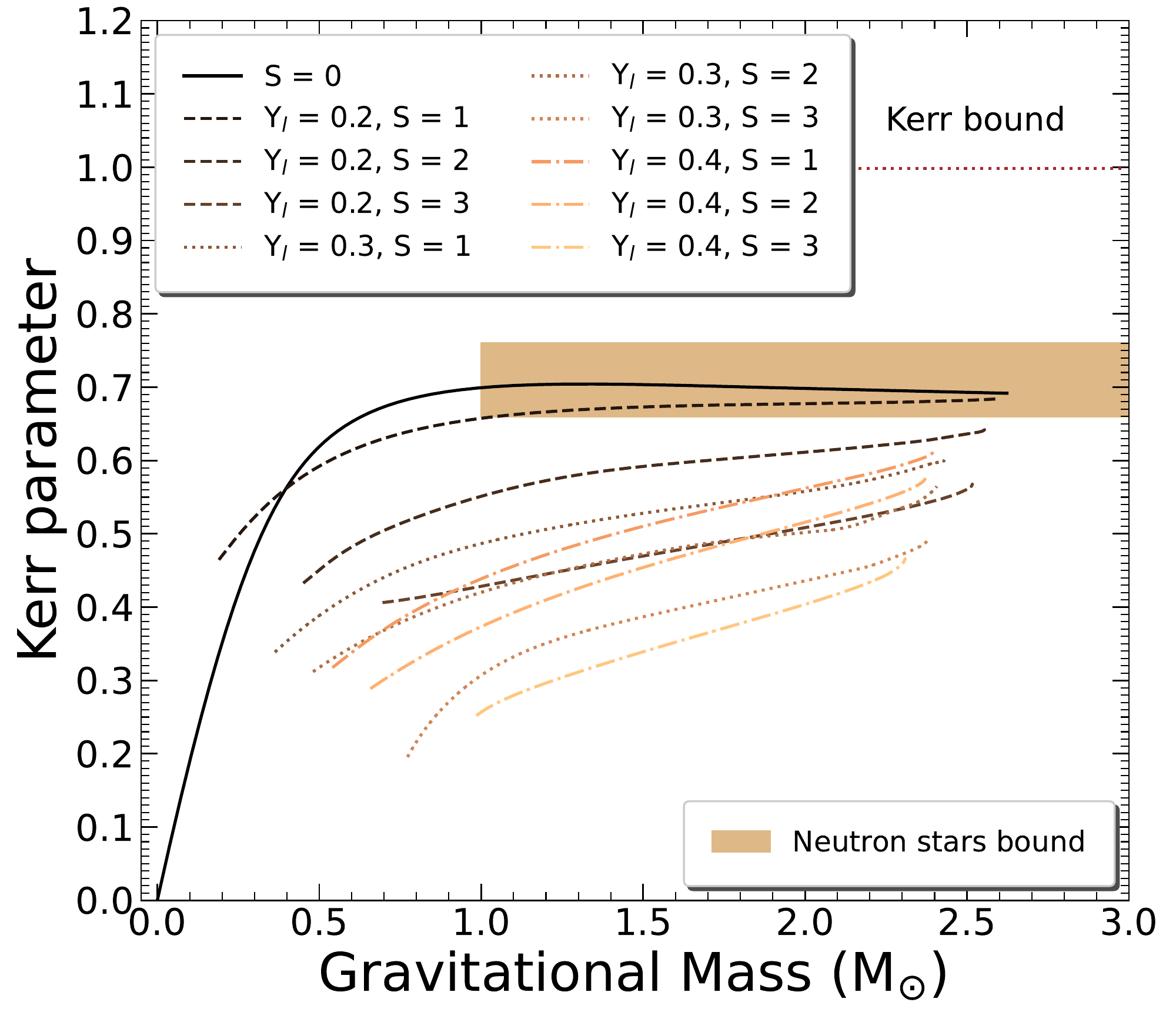}
		\caption{Kerr parameter as a function of gravitational mass for lepton fractions and entropies per baryon in the ranges $[0.2,0.4]$ and $[1,3]~k_{B}$, respectively. The horizontal dotted line marks the Kerr bound for astrophysical Kerr black holes, $\mathcal{K_{\rm B.H.}}=0.998$ (Thorne~\citeyear{Thorne_1974}). The shaded region represents the limits for neutron stars from Equation~\eqref{eq:kerr parameter 2}. The cold configuration is presented by the solid line. The configuration corresponds to the mass-shedding limit.}
		\label{fig:kerr vs mass isentropic}
	\end{figure}
	
	Figure~\ref{fig:kerr vs mass isentropic} displays the Kerr parameter as a function of the gravitational mass for isentropic neutron stars. In this scenario, the interplay between the entropy per baryon and the lepton fraction leads to different behavior for the EOS. In particular, for a constant lepton fraction, as the entropy per baryon increases, the Kerr parameter decreases.
	
	Having a limit for Kerr black holes (Thorne~\citeyear{Thorne_1974}) and one for neutron stars from Equation~\eqref{eq:kerr parameter 2} (see also Koliogiannis \& Moustakidis~\citeyear{PhysRevC.101.015805}), these values cannot be exceeded as the temperature in neutron stars increases. Therefore, the gravitational collapse of a hot, uniformly rotating neutron star, constrained to mass-energy and angular momentum conservation, cannot lead to a maximally rotating Kerr black hole. We note here that in the cold neutron star, for $M_{\rm gr} > 1~M_{\odot}$, the Kerr parameter is almost independent on the gravitational mass. However, the Kerr parameter, in the isothermal and isentropic cases, is an increasing function of the gravitational mass. This unique interplay between the angular momentum and the gravitational mass is rather significant as the temperature in the interior of the neutron star increases.
	
\begin{figure}
		\includegraphics[width=\columnwidth]{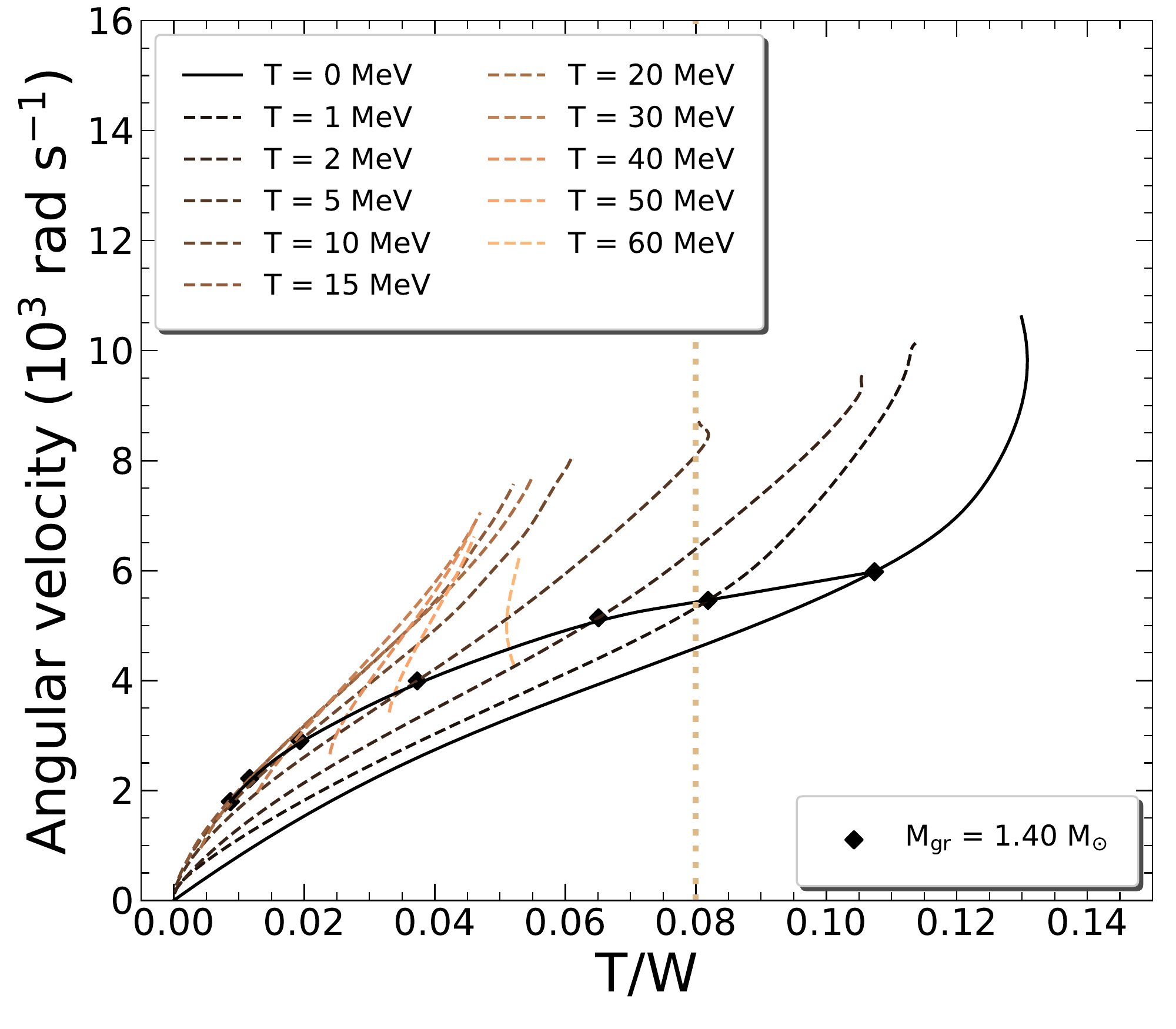}
		\caption{Angular velocity as a function of ratio of rotational kinetic to gravitational binding energy for temperatures in the range $[0,60]$ MeV. Black diamonds represent the $M_{\rm gr}=1.4~M_{\odot}$ configuration. The vertical dotted line marks the critical value, $T/W=0.08$, for gravitational radiation instabilities. The cold configuration is presented by the solid line, while the hot configurations are presented by the dashed ones. The configuration corresponds to the mass-shedding limit.}
		\label{fig:omega vs tw}
	\end{figure}

\begin{figure}
		\includegraphics[width=\columnwidth]{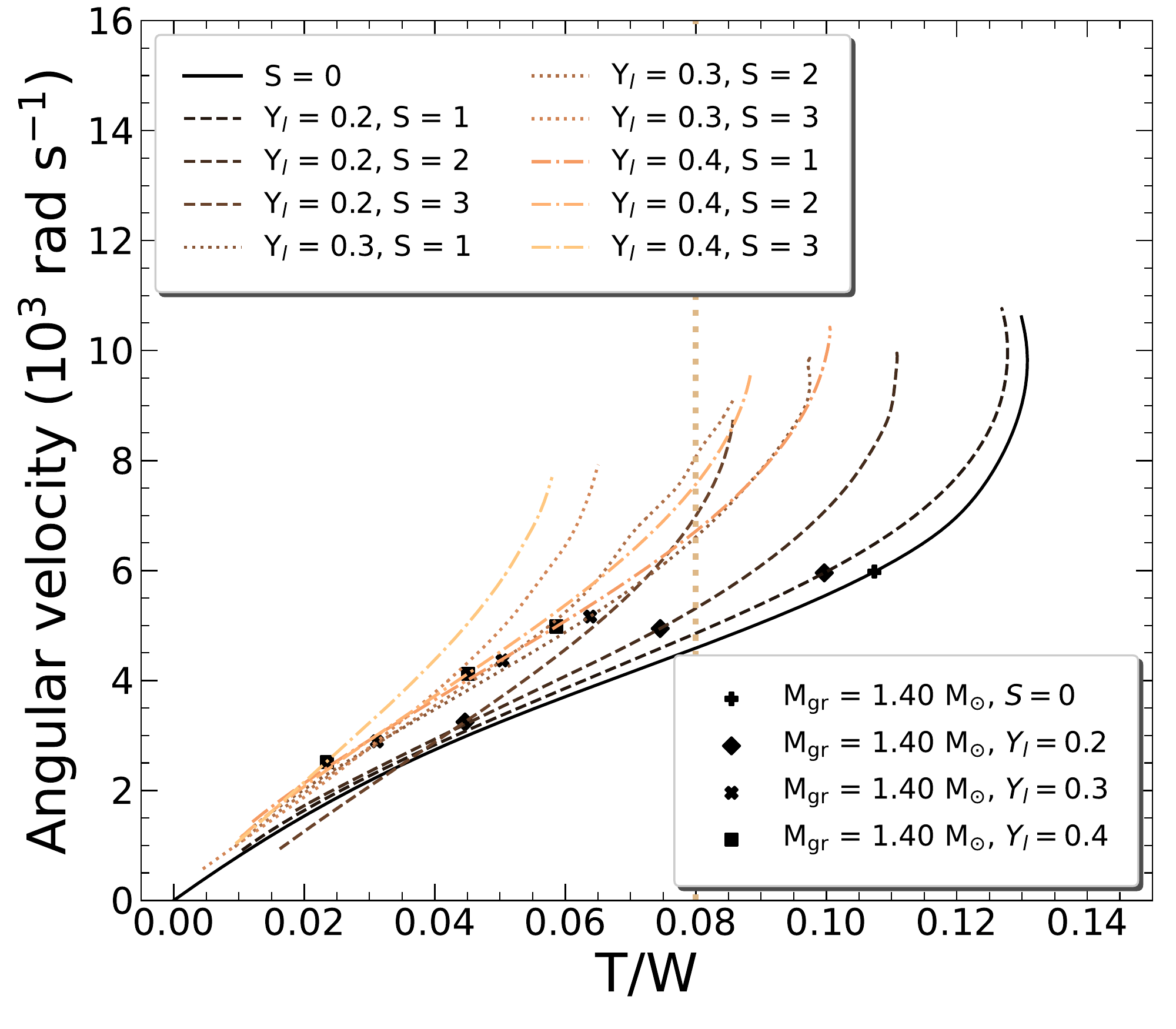}
		\caption{Angular velocity as a function of ratio of rotational kinetic to gravitational binding energy for lepton fractions and entropies per baryon in the ranges $[0.2,0.4]$ and $[1,3]~k_{B}$, respectively. Black plus sign, diamonds, squares, and crosses represent the $M_{\rm gr}=1.4~M_{\odot}$ configuration. The vertical dotted line marks the critical value, $T/W=0.08$, for gravitational radiation instabilities. The cold configuration is presented by the solid line. The configuration corresponds to the mass-shedding limit.}
		\label{fig:omega vs tw isentropic}
	\end{figure}

	Figures~\ref{fig:omega vs tw} and~\ref{fig:omega vs tw isentropic} display the angular velocity as a function of the ratio $T/W$ for isothermal and isentropic neutron stars, respectively. Nonaxisymmetric peturbations are a way for a neutron star to emit gravitational waves. In neutron stars, the point that locates the nonaxisymmetric instability is defined via the ratio of rotational kinetic to gravitational binding energy $T/W$. Instabilities driven by gravitational radiation would set in at $T/W\sim 0.08$ for models with $M_{\rm gr}=1.4~M_{\odot}$ (Morsink et al.~\citeyear{Morsink_1999}). Figures~\ref{fig:omega vs tw} and~\ref{fig:omega vs tw isentropic} show that for sufficiently compact neutron stars (EOSs with $T\leq 1~{\rm MeV}$ for isothermal and EOSs with $Y_{l}=0.2$ and $S=1$ for isentropic), the nonaxisymmetric instability will set in before the mass-shedding limit is reached. The information that can be gained is that the maximum gravitational mass, as well as the angular velocity, for a specific EOS will be lowered. Furthermore, the increasing of temperature for isothermal neutron stars leads to the conclusion that for higher temperatures than $T=2~{\rm MeV}$, the instability never occurs. In the case of isentropic ones, the increasing of entropy per baryon avoids the instability.
	
	From the relevant analysis on the quantities of this section, useful insight can be gained for the hot, rapidly rotating remnant (at least $T\geq 30~{\rm MeV}$ for isothermal EOSs, $S=1$ and $Y_{l}=0.2$ for isentropic ones) after the neutron star merger, which is a compact object with neutron star matter. The evolution of the remnant (immediately after the merger) will be one of the following four cases: (a) the one that collapses directly into a black hole, (b) the one that initially forms a neutron star but collapses during disk accretion, (c) the one that does not collapse to a black hole until after the disk has fully accreted and the newly formed neutron star spins down, and (d) the one that, even after spin-down, remains a neutron star (Fryer et al.~\citeyear{Fryer_2015}; Bernouzzi~\citeyear{Bernuzzi-2020}). In the case where the two components of the binary neutron star system have nearly the same mass, the merged object exhibits fast differential rotation. Then, depending on the strength of the magnetic field, the object quickly goes into a uniform rotation. Moreover, neutrino cooling is responsible for the redistribution of the angular momentum. This process has a very short timescale (10-100 ms; Fryer et al.~\citeyear{Fryer_2015}; Bernouzzi~\citeyear{Bernuzzi-2020}). In general, the fate of the remnant in a neutron star merger is a complicated problem, where its solution combines the use of a reliable EOS and the development of corresponding simulations. Such studies are outside the scope of the present work.
	
	Considering the maximum mass configuration at the mass-shedding limit, constraints on the hot, rapidly rotating remnant are possible through the dimensionless moment of inertia, Kerr parameter, and the ratio $T/W$. In these cases, the compactness parameter is constraint to $\beta_{\rm rem}^{\rm iso}\leq 0.19$ and $\beta_{\rm rem}^{\rm ise}\leq 0.27$, while for the Kerr parameter, the maximum allowed value is at $\mathcal{K}_{\rm rem}^{\rm iso}=0.42$ and $\mathcal{K}_{\rm rem}^{\rm ise}=0.68$ (the superscripts ``iso" and ``ise", correspond to isothermal and isentropic profiles). Concerning the ratio $T/W$, the maximum value reaches up to $(T/W)_{\rm rem}^{\rm iso}=0.05$ and $(T/W)_{\rm rem}^{\rm ise}=0.127$. Considering all of the above, two different postulations, based on isothermal and isentropic neutron stars, can be made for the aftermath of a neutron star merger. In particular, in the isothermal case, it creates a lesser compact star than the cold neutron star with lower values of maximum gravitational mass and frequency, where, for the isentropic aftermath, the object is comparable to the cold one. In addition, while in the first case, the remnant that is formed is highly stable toward the dynamical instabilities, in the second one, it is unstable.
	
	However, it has to be noted that this analysis concerns the uniform rotation. These values are expected to change if differential rotation is taken into account (Baumgarte et al.~\citeyear{Baumgarte_2000}; Morrison et al.~\citeyear{Morrison_2004}). A relevant study will be the topic of a forthcoming paper.
	
	\section{Remarks} \label{section:Remarks}
	Thermal pressure support in the isolated neutron stars and in the matter of merging (postmerger phase, remnant) is not well understood. In this study, we have attempted to gain insight into these issues by constructing and using a set of thermodynamically self-consistent EOSs (isothermal and isentropic) and also constructing nonrotating and uniformly rotating axisymmetric equilibrium sequences. Such an approximation may be acceptable for a first-order study of hot, rapidly rotating remnants of neutron star mergers, as well as protoneutron stars.
	
	The nuclear model, used in the present work, provides some advantages compared to other models, mainly that (a) the thermal effects (both in isothermal and isentropic profiles) have been included in a self-consistent way, (b) the model is flexible enough to produce EOSs from very stiff to very soft by properly modifying the density dependence of the symmetry energy, (c) the parameterization of the model is also flexible to reproduce the properties of other microscopic calculations concerning both the SNM and the PNM, (d) the momentum dependence of the potential interaction (which is absent in the majority of the proposed models) is in accordance with the terrestrial studies and experiments of heavy-ion reactions for both low and high densities and temperatures, and (e) the model ensures the causal behavior of the EOS at high densities (even at densities higher than the ones of maximum mass configuration). Future work could extend the applications on both prior and postmerger processes, including thermal effects on tidal polarizability, as well as on other bulk properties, simulations of the evolution of the merger, and processes of protoneutron stars and supernovae.
	
	The LS220 EOSs are employed for the low-density region $(n_{b}\leq 0.08~{\rm fm^{-3}})$ of hot neutron stars. For each temperature or entropy per baryon, the lowest value of the baryon density is defined at $10^{-13}~{\rm fm^{-3}}$. We found that the value of the mass is completely unaffected by the specific choice of the lowest value of the baryon density located at the surface of the star. However, as expected, the uncertainty on the value of the radius is not negligible, especially for high values of temperature (or entropy per baryon), where estimations give rise to errors at a few percent, obviously depending on the temperature (see also Raduta et al.~\citeyear{10.1093/mnras/staa2491}).
	
	Neutron stars can rotate extremely fast at the stage of being born or in the process of merging. While the Kepler frequency is an absolute limit on rotation, there are additional instabilities by which rotation may be limited if they occur at lower frequencies. However, in this study, we focus on the effect of thermal pressure. In particular, the thermal pressure begins to be less important as we reach the interior of the neutron star. However, this is not the case for the exterior region, where the bloat of the envelope takes place (Kaplan et al.~\citeyear{Kaplan_2014}). Hence, while in the case of isothermal neutron stars, hot configurations have lower frequencies than cold ones, isentropic neutron stars can possibly exceed the cold limit.
	
	The dominant quantity that manifests the thermal effects in neutron stars is the baryon mass. The baryon mass that a neutron star can support depends sensitively on the temperature, as hot neutron stars lead to lower baryon masses. Connecting this property with the merger remnant, we study the supramassive limit. In the cold case, the baryon mass is $3.085\ M_{\odot}$, while a hot one at $T=30~{\rm MeV}$ is $2.427~M_{\odot}$ and one at $S=1$ is $3.05~M_{\odot}$. These limits correspond to merger components (assuming equal masses of components) of $\sim 1.5425$, $\sim 1.2135$, and $\sim 1.525~M_{\odot}$ baryon masses, respectively. In particular, the immediate aftermath of GW170817 (Abbott et al.~\citeyear{PhysRevLett.119.161101}) and GW190425 (Abbott et al.~\citeyear{Abbott_2020}) had created a hot, rapidly rotating remnant possibly at its mass-shedding limit. Although it is most likely rotating differentially, the uniform rotation approach can provide us with useful insight about the EOS. In the case of GW170817, a remnant with a total mass of $\sim 2.7~M_{\odot}$ has been created. In correlation with the MDI+APR1 EOS, with respect to baryon mass, while the uniform rotation of cold and isentropic neutron stars can support this remnant, isothermal ones might not. Moving on to the GW190425 event, the remnant of $\sim 3.7~M_{\odot}$ cannot exist supported only by uniform rotation. However, if differential rotation is added, leading to higher masses, hot neutron stars can probably support the remnant in both cases. This possibility should be investigated further in a future work.
		
	A very recent event, GW190814 (Abbott et al.~\citeyear{Abbott_2020_a}), had a component with a mass of $\sim 2.6~M_{\odot}$. Until this moment, it was believed to be either the lightest black hole or the most massive neutron star (Most et al.~\citeyear{10.1093/mnrasl/slaa168}). However, an approach in Most et al.~\citeyearpar{10.1093/mnrasl/slaa168} suggests that this star was rapidly spinning with $\mathcal{K}$ in the range $[0.49,0.68]$. This scenario is fully supported in our study, as its mass and Kerr parameter coincide with the supramassive limit of the MDI+APR1 EOS in both cold catalyzed matter and isentropic matter with $S=1$ and $Y_{l}=0.2$. The latter may indicate that we have observed a neutron star close to or at its mass-shedding limit, being one step closer to measuring the Kepler frequency and imposing additional constraints on the EOS.
	
	Moment of inertia is a quantity that informs us about the distribution of matter in the star as it continuously changes its angular velocity and loses angular momentum due to radiation. We observed that hot neutron stars, both isothermal and isentropic ones, have lower values than the cold neutron star. This effect has its origin in the unique interplay between the gravitational mass and the equatorial radius.
	
	The Kerr parameter can be crucial as an indicator of the collapse to a black hole. Our relevant study shows that the maximum allowed value for this parameter is defined via the cold neutron star; thermal support indicates lower values of the Kerr parameter. The end point is that thermal support cannot lead a star to collapse into a maximally rotating Kerr black hole. On the other hand, the effect on the star is fascinating. Although in the cold case, after $\sim 1~M_{\odot}$, the Kerr parameter is stabilized at a constant value, when temperature is added, the Kerr parameter becomes an increasing function of the gravitational mass leading to a maximum value.
	
	The evidence related to gravitational collapse to a black hole and the existence of stable supramassive neutron stars is the ratio $T/W$. In the present study, we focus on the case of the gravitational collapse. Taking into account only the instabilities originating from gravitational radiation, the critical value of this ratio is $\sim 0.08$ for the $M_{\rm gr}=1.4~M_{\odot}$ configuration (Morsink et al.~\citeyear{Morsink_1999}). As in the case of the Kerr parameter, thermal support leads to lower values for the ratio $T/W$. As a consequence, instabilities driven by gravitational radiation never occur in a hot, rapidly rotating neutron star. However, in the specific cases of $S=1$ with $Y_{l}=0.2$ and $T< 2~{\rm MeV}$, the ratio $T/W$ deviates from the limit toward higher values. In this case, the critical value of $T/W$ may set the limit for the maximum gravitational mass and frequency. It is worth mentioning that studies related to the effect of the temperature on the Kerr parameter and the ratio $T/W$ are very rare, and their existence may open a new window in neutron star studies. 
	
	An effective way to interpret the effects of temperature on the EOS is the evolutionary sequences of constant baryon mass. From these sequences, the interest is focused on the central baryon density and its dependence on the Kepler frequency. Specifically, for temperatures $T\geq 30~{\rm MeV}$, a linear relation holds on between these quantities, leading to a universal behavior and description for the central baryon density at the mass-shedding limit. Finally, it is worth mentioning that this relation defines the allowed region of the pair of the central baryon density and corresponding Kepler frequency for a rotating hot neutron star at its mass-shedding limit.
	
	Future work should address the above analysis considering, in addition to uniform rotation, rotating configurations based on differential laws. Finally, the threshold mass and the hot, rapidly rotating remnant, as well as the possible phase transition region, should be thoroughly investigated, as the LIGO and Virgo collaboration will provide us with more events of neutron star mergers.
	
	\section{Numerical Code} \label{section: numerical code}
	The general relativistic models of neutron stars have been calculated by means of the code developed by Gourgoulhon et al.~\citeyearpar{AA.349.581}, which relies on the multidomain spectral method of Bonazzola et al.~\citeyearpar{PhysRevD.58.104020}. This code is based on the C++ library LORENE (LOREBE~\citeyear{lorene}), a software package for numerical relativity freely available under GNU license. The main characteristics of the numerical code are as follows. 
	\begin{itemize}
		\item The EOS is a barotropic one, $P=P(n)$, in a tabular form that includes the baryon density, energy density, and pressure.
		\item The whole space is divided into three domains as follows: 
		\begin{itemize}
			\item D1, the interior of the star;
			\item D2, an intermediate domain whose inner boundary is the surface of the star and outer boundary is a sphere located at $r=2r_{\rm eq}$ (where $r_{\rm eq}$ is the equatorial coordinate radius of the star); and
			\item D3, the external domain whose inner boundary is the outer boundary of D2 and that extends up to infinity.
		\end{itemize}
		\item The mapping adaptation is using one domain.
		\item The points in $\theta$, $\phi$, and $r$ are $N_{\theta} = 1\times 25$, $N_{\phi} = 1\times 1$, and $N_{r} = 3\times 49$, respectively.
		\item The initial frequency of the rotating star is $100~{\rm Hz}$ and, in low-frequency areas ($< 100~{\rm Hz}$), $10/50~{\rm Hz}$.
		\item The global numerical error is evaluated by means of the virial identities, GRV2 and GRV3, where the latter is a relativistic generalization of the classical virial theorem. For the configurations presented in this paper, the relative errors are of order $10^{-6}$.
	\end{itemize}
	
	\begin{acknowledgments}
	The authors thank Prof. K. Kokkotas for his constructive comments on the preparation of the manuscript and Profs. D. Radice and N. Stergioulas for the useful correspondence.
	We also thank Prof. L. Rezzolla for his useful and helpful considerations and clarifications.
	\end{acknowledgments}

	\software{\emph{nrotstar} from C++ Lorene/Nrotstar library (LORENE~\citeyear{lorene})}

	\appendix
	\section{Properties of Nuclear Matter}
	\label{section:appendix_1}
	The total energy per particle can be expanded as follows:
	\begin{align}
		E(n,I) =& E(n,0)+\sum_{k=2,4,\cdots}E_{{\rm sym},k}(n)I^{k}, \quad \text{where} \nonumber \\ E_{{\rm sym},k}(n)=&\left. \frac{1}{k!}\frac{\partial^{k} E(n,I)}{\partial I^{k}}\right|_{I=0}.
		\label{Expand-1}
	\end{align}
	
	In particular, we considered the PA in which the symmetry energy is given through
	\begin{equation}
		E_{\rm sym}(n) = E(n,I = 1) - E(n,I = 0).
		\label{PA-1}
	\end{equation}
	
	The properties of nuclear matter at the saturation density are defined as (Costantinou et al.~\citeyear{PhysRevC.89.065802}, ~\citeyear{PhysRevC.92.025801})
	\begin{align}
		L =& 3n_s\left. \frac{dE_{\rm sym}(n)}{dn}\right|_{n_s}, \quad K = \left.9n_{s}^{2} \frac{d^{2} E_{\rm sym}(n)}{dn^{2}}\right|_{n_{s}}, \nonumber \\ Q =& 27n_s^3\left. \frac{d^{3} E_{\rm sym}(n)}{dn^{3}}\right|_{n_{s}},
		\label{K-1}
	\end{align}
	
	\begin{equation}
		K_{s} = \left.9n_s^2\frac{d^2 E(n,0)}{dn^2}\right|_{n_s}, \quad Q_{s} = 27n_{s}^{3}\left. \frac{d^{3} E(n,0)}{dn^{3}}\right|_{n_{s}},
		\label{L-1}
	\end{equation}
	where $L$, $K$, and $Q$ are related to the first, second, and third derivative of the symmetry energy $E_{\rm sym}(n)$, respectively. Here $K_{s}$ is the compression modulus, and $Q_{s}$ is related to the third derivative of $E(n,0)$. The last property is the ratio of the Landau effective mass to mass in vacuum for the MDI model (Prakash et al.~\citeyear{PRAKASH19971}; Moustakidis~\citeyear{PhysRevC.89.065802}; Constantinouu et al.~\citeyear{PhysRevC.92.025801}, ~\citeyear{PhysRevC.78.054323}) and given by
	\begin{equation}
		\frac{m^{*}_{\tau}(n,I)}{m_{\tau}} = \left[ 1 - \frac{2nm_{\tau}}{n_{s}\hbar^{2}} \sum_{i=1,2} \frac{1}{\Lambda_{i}^{2}}\frac{C_{i} \pm \frac{C_{i}-8Z_{i}}{5}I}{\left[1 + \left(\frac{k_{F}^{0}}{\Lambda_{i}}\right)^{2} \left[\left(1 \pm I\right)\frac{n}{n_{s}}\right]^{2/3}\right]^{2}}\right]^{-1}.
		\label{eq:lem}
	\end{equation}
	where $\tau$ corresponds to neutrons or protons.
	
	\bibliography{koliogiannis}{}
	\bibliographystyle{aasjournal}

\end{document}